\documentclass[twocolumn]{aa}
\usepackage{float}
\usepackage{graphicx}

\usepackage{lineno}
\usepackage{multirow}
\usepackage{subfigure}
\usepackage[]{xcolor}
\definecolor{citecolor}{HTML}{195D95}
\definecolor{green}{HTML}{20C29D}
\definecolor{BrickRed}{HTML}{C41010}

\usepackage[colorlinks=true,linkcolor=green,citecolor=citecolor,urlcolor=green]{hyperref}
\usepackage{amsmath,amssymb,latexsym}

\def\vFv{\nu F_{\nu}}

\def\gcool{\gamma_{\rm cool}}
\def\gmin{\gamma_{\rm inj}}
\def\gmax{\gamma_{\rm max}}

\def\hnuc{h \nu_{\rm cool}}
\def\hnumax{h \nu_{\rm max}}

\def\ngamma{n_{\gamma}}
\def\Rgamma{R_{\gamma}}
\def\ntheta{n_{\theta}}
\def\btheta{b_{\theta}}
\def\Rtheta{R_{\theta}}
\def\pfrac{\bar{p}}

\def\npha{n_{\mathrm{pha}}}
\def\Rgamma{R_{\gamma}}
\def\ntheta{n_{\theta}}
\def\btheta{b_{\theta}}
\def\Rtheta{R_{\theta}}
\def\pfrac{\bar{p}}

\newcommand{\cond}[3]{\pi_{\tiny{#1}}\left(#2 \;|\; #3 \right)}

\def\diff#1{\mathrm{d}#1 \;}
\def\ene{\varepsilon}

\begin{document}

\title{Time-resolved GRB polarization with POLAR and GBM}
\subtitle{Simultaneous spectral and polarization analysis with synchrotron emission}

\author{J. Michael Burgess \inst{1,2} 
  \and M. Kole\inst{3}
  \and F. Berlato \inst{1}
  \and J. Greiner \inst{1,2}
  \and G. Vianello \inst{4}
  \and N. Produit \inst{5}
  \and Z.H Li \inst{6,7}
  \and J.C Sun \inst{6}
}

\institute{Max-Planck-Institut fur extraterrestrische Physik, Giessenbachstrasse 1, D-85748 Garching, Germany \\
  \email{jburgess/fberlato/jcg@mpe.mpg.de}\label{mpe}
  \and Excellence Cluster Universe, Technische Universit{\"a}t M{\"u}nchen, Boltzmannstra{\ss}e 2, 85748 Garching, Germany\label{ec}
\and University of Geneva (DPNC), quai Ernest-Ansermet 24, 1205 Geneva, Switzerland \\ \email{merlin.kole@unige.ch}\label{dpnc}
\and Hansen Experimental Physics Lab, Stanford University, 452 Lomita Mall, Stanford, CA 94305-4085,
USA\label{stanford}
\and Astronomy Departement University of Geneva, Versoix, Switzerland
\and Key Laboratory of Particle Astrophysics, Institute of High Energy Physics, Chinese Academy of Sciences, Beijing 100049, China
\and University of Chinese Academy of Sciences, Beijing 100049, China
}
%\linenumbers
\date{}

\label{firstpage}

\abstract {Simultaneous $\gamma$-ray measurements of $\gamma$-ray burst
  spectra and polarization offer a unique way to determine the
  underlying emission mechanism(s) in these objects, as well as probing
  the particle acceleration mechanism(s) that lead to the observed
  $\gamma$-ray emission} {We examine the jointly observed data
  from POLAR and \textit{Fermi}-GBM of GRB 170114A to determine its
  spectral and polarization properties, and seek to understand the
  emission processes that generate these observations. We aim to
  develop an extensible and statistically sound framework for these
  types of measurements applicable to other instruments.} {We leveraged
  the existing \texttt{3ML} analysis framework to develop a new
  analysis pipeline for simultaneously modeling the spectral and
  polarization data. We derived the proper Poisson likelihood for
  $\gamma$-ray polarization measurements in the presence of
  background. The developed framework is publicly available for
  similar measurements with other $\gamma$-ray polarimeters. The data
  are analyzed within a Bayesian probabilistic context and the
  spectral data from both instruments are simultaneously modeled with
  a physical, numerical synchrotron code.} {The spectral modeling of
  the data is consistent with a synchrotron photon model as has been
  found in a majority of similarly analyzed single-pulse gamma-ray
  bursts. The polarization results reveal a slight trend of growing
  polarization in time reaching values of $\sim 30$\% at the temporal
  peak of the emission. We also observed that the
  polarization angle evolves with time throughout the emission. These
  results suggest a synchrotron origin of the emission but further
  observations of many GRBs are required to verify these evolutionary
  trends. Furthermore, we encourage the development of time-resolved
  polarization models for the prompt emission of gamma-ray bursts as
  the current models are not predictive enough to enable a full
  modeling of our current data.} {}


\maketitle

\section{Introduction}
Polarization measurements from astrophysical objects are a key piece
of information to decipher the physics and geometry of regions that
emit the observed photons. The emission from $\gamma$-ray bursts
(GRBs) has been notoriously difficult to understand due to the
complexity of modeling their broadband, prompt $\gamma$-ray
emission. Recent results have provided evidence that the prompt
emission is the result of synchrotron radiation from electrons
accelerated to ultra-high energies via magnetic reconnection
\citep{Burgess:2014aa,Zhang:2016aa,Zhang:2018aa,Burgess:2018}. Measurements
of the optical polarization from a GRB's prompt emission have
similarly pointed to a synchrotron origin of the emission
\citep{Troja:2017aa}. However, spectral modeling of photospheric based
emission has also provided adequate fits to a subset of GRBs
\citep{Ryde:2010aa,Ahlgren:2015aa,Vianello:2018il}. Measurements of
polarization can break this degeneracy
\citep{Toma2009,Gill:2018aa}. Photospheric emission will typically
produce unpolarized emission although a moderate polarization level is
possible in special circumstances \citep{Lundman:2018aa} and predicts
very specific changes of the polarization angle
\citep{Lundman:2014aa}. On the other hand, synchrotron emission
naturally produces a range of polarized emission depending on the
structure of the magnetic field and outflow geometry
\citep{Waxman2003,Lyutikov:2003, JG2003}. Thus, being able to fit
synchrotron emission to the observed spectrum while simultaneously
detecting polarization provides a clear view of the true emission
process.

Several reports of polarization measurements have been produced by a
variety of instruments. An overview of which can be found in
\citep{Covino2016}. Of these measurements, those by non-dedicated
instruments like those reported by BATSE and RHESSI suffer from
problems with instrumental effects or poorly understood systematics
\citep{MCCONNELL20171} making it impossible to draw conclusions based
on these results. Additionally, several measurements were performed
using data from two instruments onboard the INTEGRAL satellite, IBIS,
and SPI. Several of the GRB polarization measurements performed by
these instruments do not suffer from obvious errors in the analysis
and allow us to constrain the polarization parameter space. However,
for several of these measurements systematic uncertainties also make
it difficult to draw conclusions \citep{McGlynn2007}. Furthermore as
stated in for example \citep{PEARCE201954}, a lack of on-ground
calibration of the instrument responses of both IBIS and SPI to
polarized beams creates additional doubt on the validity of
polarization results from these instruments within the community. This
indicates the importance of performing polarization measurements with
carefully calibrated and dedicated instrumentation.  More recently,
the AstroSAT collaboration has reported preliminary polarization
analysis results of several GRBs on the arXiv e-Print archive
\citep{Chattopadhyay:2017aa}. The systematics and
procedures related to obtaining these measurements is not
immediately clear. The quoted error distributions contain unphysical
regions of parameter space (polarization degrees greater than 100\%)
and are thus questionable. Past measurements of polarization by the
first dedicated GRB polarimeter, GAP, provided hints of polarized
emission \citep{Yonetoku:2011ef}. The results presented there indicate
an overall low polarization potentially resulting from an evolution of
the polarization angle during the long multipulse GRB, something also
reported in \citep{G_tz_2009} for GRB 041219A. Measurements by COSI
provided an upper limit on the polarization degree
\citep{Lowell:2017bq}. The statistics of these measurements do not,
however, allow constraints on the emission mechanisms. Furthermore,
the techniques for all these measurements relied on background
subtraction. As both the background and signal counts are Poisson
distributed, subtraction is an invalid procedure that destroys
statistical information, thus all reported significances are
questionable.

The POLAR experiment \citep{Produit2018} on board the Chinese space
laboratory Tiangong-2 observed 55 GRBs and reported polarization
measurements for five of these GRBs \citep{POLAR2018}. Time-integrated
analysis of these GRBs resulted in strict upper limits on the
polarization degrees. The most likely polarization degrees found in
that analysis are non-zero but remain compatible with an unpolarized
emission, leading to the conclusion that GRBs are at most moderately
polarized. Using time-resolved analysis it was however found that the
polarization of GRB 170114A was most compatible with a constant
polarization degree of $\sim 28\%$ with a varying polarization
angle. Summing polarized fluxes with varying polarization degrees
produces an unpolarized flux. The detection of an evolution in
polarization angle within this single pulse GRB could explain the low
polarization degrees found for all five GRBs. The results presented in
\citep{POLAR2018} do not, however allow for a detailed time-resolved
study of the remaining four GRBs, nor do they allow determination of
the nature of the evolution of the polarization angle in GRB 170114A.

Coincidentally, several of the GRBs observed by POLAR were
simultaneously observed by the \textit{Fermi}-GBM. In this paper, we present
a technically advanced modeling of the polarization and spectral data
simultaneously with data from both instruments. This allows the
incorporation of information contained in both data sets leading to
improved sensitivity and an altogether more robust analysis. This work
is organized as follows: The methodology and modeling is described in
Sections \ref{sec:method} and \ref{sec:synch} and the results are
interpreted in Section \ref{sec:results}.

\section{Data analysis and methodology}
\label{sec:method}
For the analysis herein, we have developed a new approach of
simultaneously fitting both the spectral data from POLAR and GBM along
with the POLAR scattering angle (SA) or polarization data
(the subset of POLAR data usable for polarization analysis
  selected with cuts as defined in \citep{LI2018}). This simultaneous
fitting alleviates the need for approximate error propagation of the
spectral fits into the polarization analysis. Using the abstract data
modeling capabilities of
\texttt{3ML}\footnote{\url{https://threeml.readthedocs.io/en/latest/}}
\citep{Vianello:2015}, a framework was developed to directly model all
data simultaneously with a joint-likelihood in each dataset's
appropriate space. Below, we describe in detail  each part of the methodology.

We focus on the analysis of GRB 170114A \citep{Veres:2017}
  which is a single-pulse, bright GRB lasting approximately 10s which
  allows us to performed detailed time-resolved spectroscopy. The
  event occurred on January 14$^{\text{th}}$ 2017 with an initially
  estimated fluence between 10-1000 keV of $\sim 1.93 \cdot 10^{-5}$
  erg cm$^{-2}$. The high peak flux of the GRB triggered an
  autonomous repoint request for the $Fermi$ satellite, however, no
  LAT detection of photons occurred.

\subsection{Location and temporal analysis}
\begin{figure}
  \centering
 
  \includegraphics{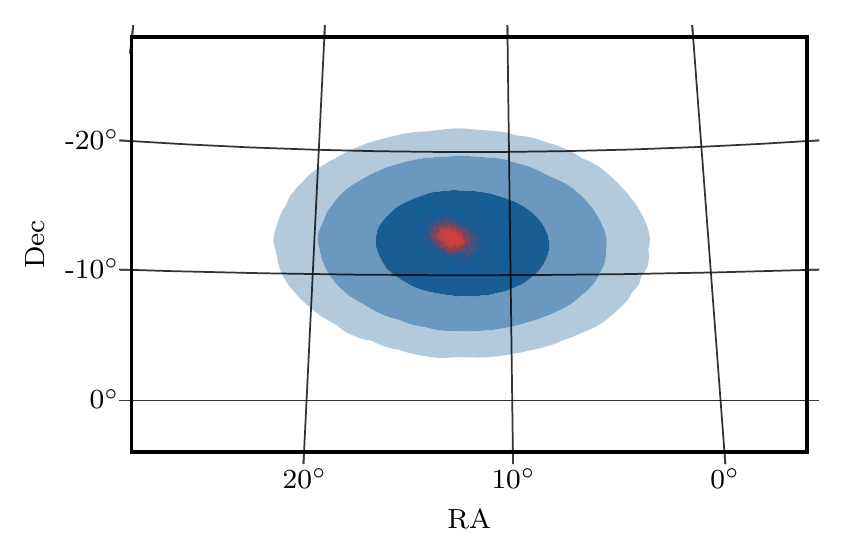}
  \caption{BALROG location (red posterior samples) of GRB 170114A
    derived by fitting the peak of the emission for both the location
    and spectrum simultaneously. The blue contours display the 1,2,
    and 3 $\sigma$ standard GBM catalog location as obtained from the
    Fermi Science Support Center (FSSC).}
  \label{fig:location}
\end{figure}

Spectral and polarization analysis for both GBM and POLAR rely on
knowledge of the sky-position ($\delta$) of the GRB in question. As they are both all-sky surveyors, GBM and POLAR lack the ability to image GRBs
directly. However, using the BALROG technique \citep{Burgess:2018et},
we can use the spectral information obtained in the GBM data to locate
the GRB. Using a synchrotron photon model (see Section
\ref{sec:synch}), we were able to locate the GRB to RA=
$13.10 \pm 0.5$ deg, DEC=$-13.0 \pm 0.6 $ deg. Using this location,
spectral and polarization responses were generated for all data
types. We note that a standard GBM position \footnote{Data obtained
  from \url{https://heasarc.gsfc.nasa.gov/FTP/fermi/data/gbm/bursts/}}
exists and, along with their uncertainties, was used for the
polarization results presented in \citep{POLAR2018}, however, the
standard localization technique has known systematics and now possess
arbitrarily inflated error distributions
\citep{Connaughton:2015aa}. We find the BALROG derived location much
more precise than that of the standard location analysis (see Fig
\ref{fig:location}), allowing us to reduce the systematic errors included
in the polarization results presented in
\citep{POLAR2018}. Additionally, it has now been shown that the BALROG
locations as systematically more accurate \citep{Berlato:2019}.

\begin{figure*}%
  \centering
\begin{subfigure}[]%
{\includegraphics[width=3.4in]{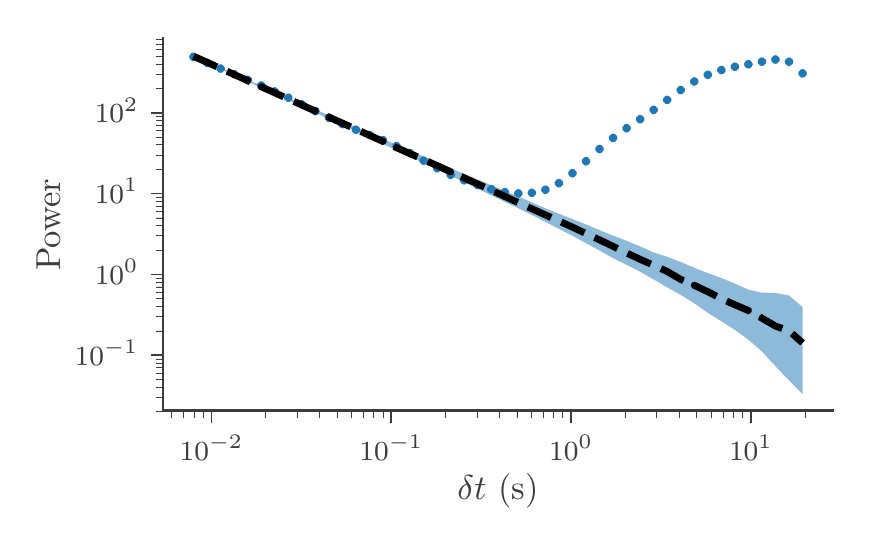}}%
\end{subfigure}%
\begin{subfigure}[]%
{\includegraphics[width=3.4in]{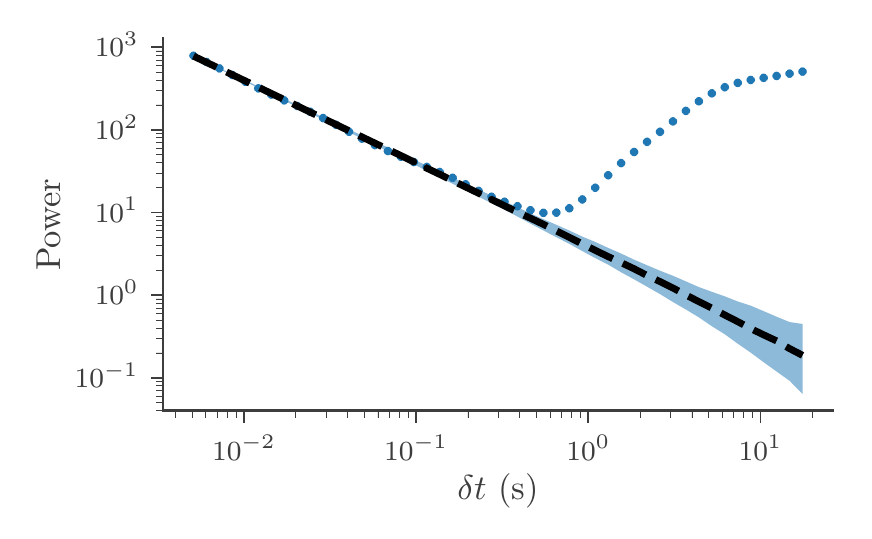}}%
\end{subfigure}%
\caption{Minimum variability timescales for the polar
  polarization data (left) and the GBM spectral data (right). The
  black line indicates the background power spectrum determined via
  Monte Carlo calculations and the shaded regions indicates the
  uncertainty in the background. Notably, both data sets have nearly
  equivalent MVTs.}
\label{fig:mvt}
\end{figure*}

The chief focus of this analysis is temporal variation in the
polarization parameters. We computed the minimum variability timescale
(MVT) \citep[see][ for details]{Vianello:2018il} on the POLAR SA light
curve. The MVT infers the minimum timescale above the
  Poisson noise floor of which variability exists in the data. This
yields an MVT of $\sim0.3$ s (Fig. \ref{fig:mvt}). For completeness,
the MVTs for both the GBM and POLAR spectral light curves were
computed as well. Both analyses yield similar results. Therefore, we
were able to analyze data on this timescale without the concern of
summing over evolution of spectral \citep{Burgess:2015iy}. However,
the raw polarization data do not allow for us to check for variability
in the polarization angle prior to fitting. Therefore, it is possible
that the angle could change on a timescale smaller than our selected
time-intervals. This could reduce the overall inferred polarization.

With the MVT determined, we utilized the Bayesian blocks algorithm
\citep{Scargle:2013aa} to objectively identify temporal bins for the
analysis. The SA light curve was utilized to perform the analysis. The
temporal bins created are on the order of the MVT. A total of nine
bins were selected and used for spectral and polarization analysis
(see Table \ref{tab:table1}).

\begin{figure}
  \centering
  \includegraphics{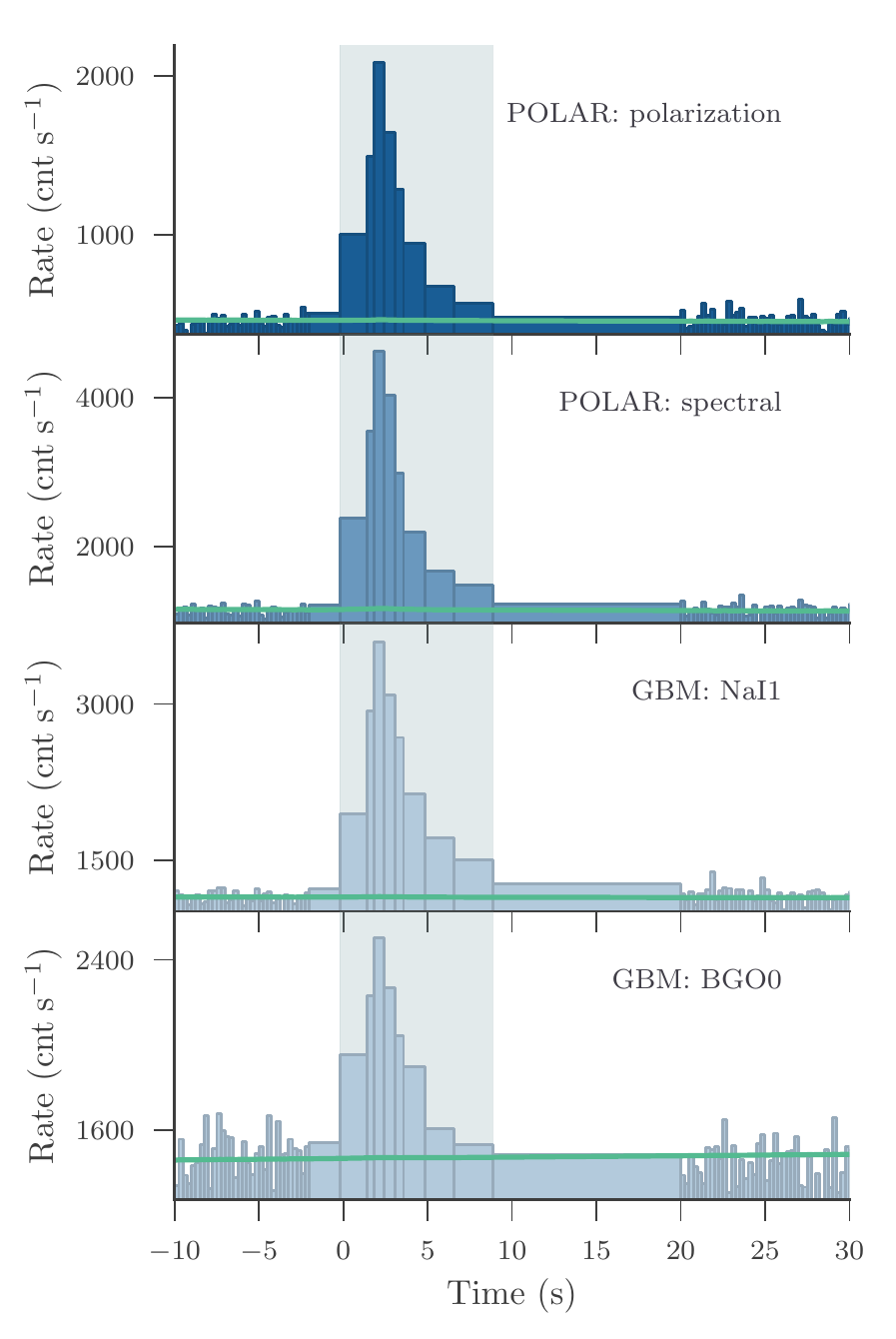}
  \caption{Light curves of the POLAR polarization and spectral
    data (the difference is explained in appendix
    \ref{sec:polar_response}) as well as two GBM detector data. The
    green line is the fitted background model and the gray shaded
    regions show the time-intervals used for the analysis. The
    binning in the analysis region is derived via Bayesian
    blocks. }
  \label{fig:lightcurve}
\end{figure}

\subsection{Spectral analysis}
The standard $\gamma$-ray forward-folding approach to spectral fitting
is adopted, in which we have sky location ($\delta$) dependent responses
for both the GBM and POLAR detectors ($\Rgamma$) and fold the proposed
photon model ($\ngamma$) solution through these responses to produce
detector count spectra ($\npha$). Thus,

\begin{equation}
  \label{eq:2} \npha^{i,j} = \int \diff{\ene^j} \ngamma(\ene,\bar{\psi}) \Rgamma^{i,j} \left(\delta \right)
\end{equation}
\noindent
for the $i^{\mathrm{th}}$ detector in the $j^{\mathrm{th}}$
pulse-height amplitude (PHA) channel, $\ene$ is the latent photon
energy and $\bar{\psi}$ are a set of photon model parameters. Here,
$\delta$ is the sky location of the GRB. Both POLAR and GBM have Poisson-distributed
total observed counts, and their backgrounds
determined via fitting polynomials in time to off-source regions of
the light curves. Thus, Gaussian-distributed background counts are
estimated by integrating these polynomial models over the source
interval of interest. The uncertainty on these estimated counts is
derived via standard Gaussian uncertainty propagation. This leads us
to use a Poisson-Gaussian likelihood\footnote{This is known as PGSTAT
  in XSPEC.} for each detector for the spectral fitting.

\subsection{Polarization analysis}
\begin{figure}

  \centering
\includegraphics[]{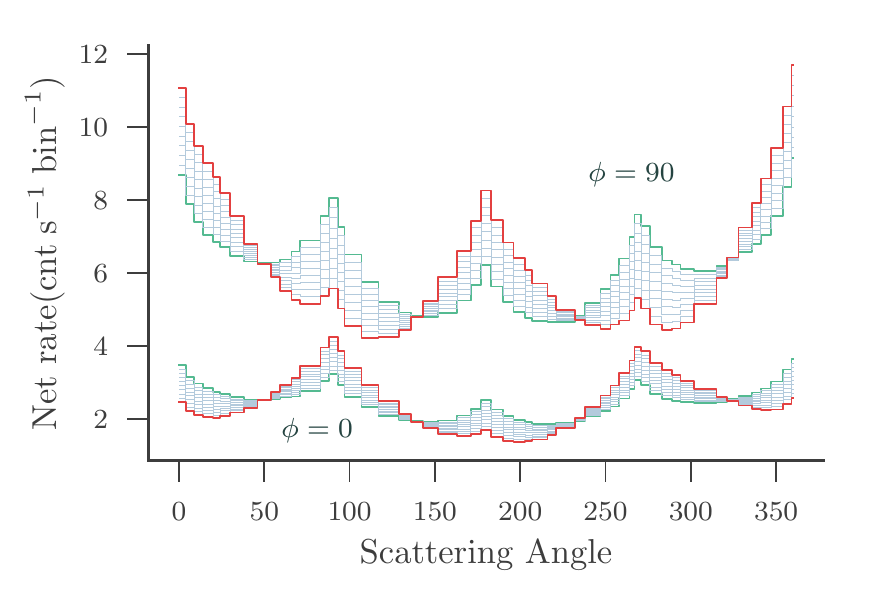}
\caption[]{Folded POLAR count space for two polarization angles
  and ten levels of polarization degree. The rates have been
  artificially scaled to between the different angles for visual
  clarity. The green lines for both angles represent the polarization
  degree $\bar{p}=0,$ and the red lines $\bar{p}=100$. Thus, we can see how various sets of polarization parameters can be identified
  in the data.  The peaks with a $90^\circ$ periodicity are the result
  of POLAR's square shape, while the visible modulation with a
  $360^\circ$ period is a result of the incoming direction of the
  photons with respect to the instrument's zenith. By forward-modeling
  the instrument response, the systematics induced by geometrical
  effects are properly accounted for.}
  \label{fig:demo}
\end{figure}

To enable performing joint fits of the spectra and the polarization a
novel analysis technique was developed. Traditional polarization
analysis techniques, such as those employed in
\citep{Yonetoku:2011ef,Chattopadhyay:2017aa} as well as in
\citep{POLAR2018}, rely on fitting data to responses produced
for a specific spectrum. This method does not allow joint fits of both
the spectrum and polarization parameters, nor does it allow naturally
including systematic uncertainties from the spectral fits into the
systematic uncertainties of the polarization. Here, in order to model
the polarization signal seen in the data, we invoked a forward-folding
method similar in concept to our approach to spectral analysis. We
simulated polarized signals as function of polarization angle, degree
and energy to create a matrix of SA distributions (often
called modulation curves within the field of polarimetry) which can be
compared to the data via the likelihood in data space. For details on
the creation of the matrix see Appendix \ref{sec:polar_response}.
Mathematically,

\begin{equation}
  \label{eq:1} \ntheta^{k} \left(\phi, \pfrac \right) = \int \diff{\ene^{j}} \ngamma \left(\ene; \bar{\psi} \right) \Rtheta^{j,k} \left(\ene, \phi, \pfrac \right)
\end{equation}
\noindent
where $\ntheta$ are counts in SA bin $k$, and $\Rtheta$
is the simulated response of the corresponding scattering bin. In
words, we convolved the photon spectrum over the $j^{\mathrm{th}}$
photon energy bin with the polarization response to properly weight
the number of counts observed in each SA bin. Figure
\ref{fig:demo} demonstrates how changes in polarization angle and
degree appear in the POLAR data space. Hence, our need to
simultaneously fit for the photon spectrum which allows for direct
accounting of the uncertainties in the weighting.

POLAR observed SAs are measured as detector counts and thus Poisson
distributed. The pollution of the source signal by background cannot
be handled by background subtraction as has been done in previous
work. Instead, a temporally off-source measurement of the background
polarization is made in order to model the background contribution to
the total measurement during the observation intervals. The background
measurement is Poisson distributed in each of the $k$ scattering
bins. Due to the temporal stability of the background, as presented in
\citep{POLAR2018}, we fit a polynomial in time to each of the $k$
scattering bins via an unbinned Poisson likelihood. This allowed us to
reduce the uncertainty of the background by leveraging the temporal
information. We were able to estimate the on-source background contribution
($\btheta^k$) by integrating the polynomials over time and propagating
the temporal fit errors. This implies that the polarization likelihood
is also a Poisson-Gaussian likelihood just as with the spectral
data. We verified that our approach allowed us to identify the latent
polarization parameters via simulations in Appendix
\ref{sec:polar_assessment}. The count rates are corrected for the
proper exposure by computing the total dead-time fraction associated
with each interval. The method employed for dead-time calculation is
equivalent to that of \citet{POLAR2018}.

\begin{figure}
  \centering
  \includegraphics{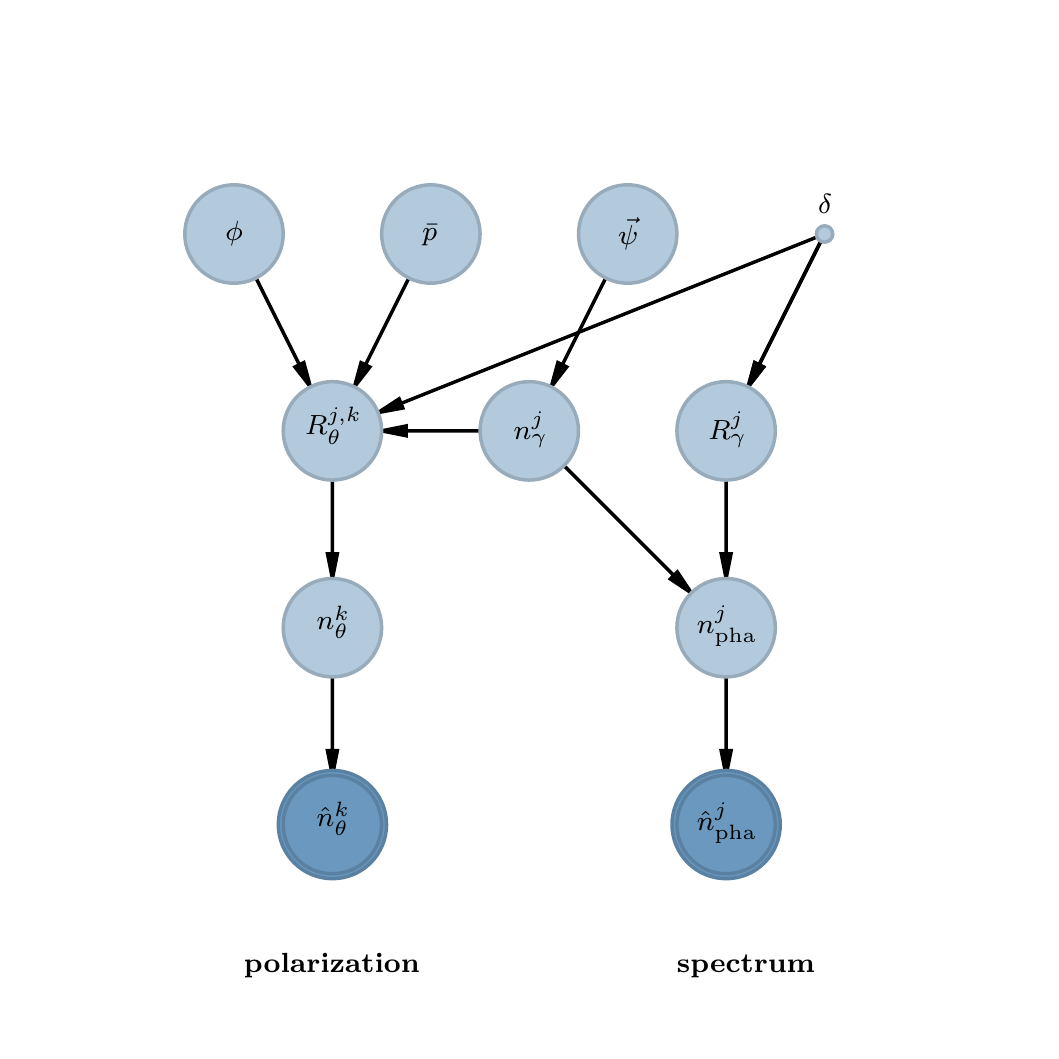}
  \caption{Directed graph model describing the full likelihood of
    our approach. Model parameters are shown in light blue, and the data in
    dark blue. The graph shows how the latent parameters of the model
    are connected to each other and eventually the data. It is
    important to note that the latent photon model connects both sides
    of the model. The position ($\delta$) is a fixed parameter. Here
    $\vec{\psi}$ represents the set of spectral parameters.  }
  \label{fig:like}
\end{figure}

The full joint likelihood of the data is thus a product over the
spectral and polarization likelihoods which is detailed in Appendix
\ref{sec:likelihood} (see also Figure \ref{fig:like}). We re-emphasize
that the spectral model and polarization model communicate with each
other through the likelihood. This implies that the posterior density
of the model is fully propagated to both datasets without any
assumptions such as Gaussian error propagation. As is seen in the
following sections, the resulting parameter distributions can be
highly asymmetric.

In a perfect world where all instruments are cross-calibrated over the
full energy range, the instruments' various responses would predict
similar observed fluxes for each measurement. However, we allowed for a
normalization constant between GBM and POLAR to account for any
unmodeled discrepancies between the instruments. Both POLAR's
polarization and spectral data are scaled by these constants which are
unity when no correction is required\footnote{We could have easily
  applied these constants to the GBM responses. Since they are
  scalings, where they are applied is arbitrary.}. This constant scale
for the effective area by no means accounts for energy-dependent
calibration issues.

In order to obtain the posterior parameter distributions, we used {\tt
  MULTINEST} \citep{Feroz:2009aa, Buchner:2014aa} to simulate the model's
posterior. {\tt MULTINEST} utilizes nested sampling which is suitable
for the multimodal distributions we observe, as well as for the
non-linear model and high-dimension of our parameter space. For the
polarization parameters, we used uninformative priors of appropriate
scale. The effective area normalizations are given informative
(truncated Gaussians) priors centered at unity with a 10\% width. The
priors for the spectral modeling are discussed in the Section
\ref{sec:synch}. We ran {\tt MULTINEST} with 1500 live points to
achieve a high number of samples for posterior inference. Model
comparison was not attempted and thus we did not use the marginal
likelihood calculations\footnote{Indeed, astrophysical models operate
  in the $\mathcal{M}$-open probabilistic setting and marginal
  likelihood is an $\mathcal{M}$-closed tool \citep{Vehtari:2018aa}.}.

As stated, for both $\bar{p}$ and $\phi$, we used uninformative priors
in each parameters' domain. This is a valid choice for $\phi$, but we
note that an informative prior for the expected polarization from
synchrotron emission could be used as an assumption. However, as
discussed in Section \ref{sec:discussion}, the theoretical predictions
for GRB synchrotron models are not mature enough for us to assume such
a prior at the current time. Nevertheless, in our work we tested
Gaussian priors centered at moderate polarization and found that the
data allowed for this assumption. Moreover, we found that our
recovered $\phi$ was not affected by out choice of prior on $\bar{p}$.

\section{Synchrotron modeling}
\label{sec:synch}
With the recent finding that synchrotron emission can explain the
majority of single-pulse GRBs, we chose to model the time-resolved
photon spectrum with a physical synchrotron model. Following
\citet{Burgess:2018}, we set

\begin{equation}
  \label{eq:3}
    \ngamma \left(\ene ; K, B, p, \gcool \right) = \int_{0}^{t^{\prime}(\gcool)} \int_{1}^{\gmax} \diff{t} \diff{\gamma} \times n_e \left(\gamma; t \right) \Phi\left(\frac{\ene}{\ene_{\mathrm{crit}}(\gamma; B ) } \right)
,\end{equation}
\noindent
where $K$ is the arbitrary normalization of the flux, $B$ is the
magnetic field strength in Gauss, $p$ is the injection index of the
electrons, $\gcool$ is the energy to which an electron will cool
during a synchrotron cooling time,
\begin{equation}
  \label{eq:4}
  \Phi\left( w\right) = \int_{w}^{\infty} \diff{x} K_{5/3} \left(x \right)
\end{equation}
\noindent
and
\begin{equation}
  \label{eq:5}
  \ene_{\mathrm{crit}} \left(\gamma ; B \right) = \frac{3}{2} \frac{B}{B_{\mathrm{crit}}} \gamma^2 \mathrm{.}
\end{equation}
\noindent
Here, $K_{5/3}$ is a Bessel function,
$B_{\mathrm{crit}} = 4.13 \cdot 10^{13}$ G, and $n_e$ is determined by
solving the cooling equation for electrons with the Chang and Cooper
method \citep{Chang:1970gk}. In mathematical expression,
\begin{equation}
  \label{eq:6}
  \frac{\partial}{\partial t} n_e \left(\gamma, t \right) = \frac{\partial}{\partial t} \dot{\gamma}\left( \gamma; B \right) n_e \left(\gamma, t \right) + Q(\gamma;\gmin, \gmax, p)
,\end{equation}
\noindent
where the injected electrons are defined by a power law of index $p$
\begin{equation}
  \label{eq:7}
  Q\left(\gamma; \gmin, \gmax ,p  \right) \propto \gamma^{-p}\;\; \gmin \le \gamma \le \gmax
,\end{equation}
\noindent
where $\gmin$ and $\gmax$ are the minimum and maximum injected
electron energies respectively and the synchrotron cooling is
\begin{equation}
  \label{eq:8}
  \dot{\gamma}\left( \gamma ; B\right) = -\frac{\sigma_{\mathrm{T}} B^2 }{6 \pi m_e c} \gamma^2 \mathrm{.}
\end{equation}
\noindent

For our numerical calculations we created a 300-point grid,
logarithmically distributed in $\gamma$. The linear equations in the
implicit scheme form a tridiagonal matrix which is solved numerically
with standard methods. The method of \citet{Chang:1970gk} is
numerically stable and inexpensive as well as shown to conserve
particle number in the absence of sources and sinks. Thus, we are able
to solve for the synchrotron emission spectrum quickly during each
iteration of the fit. The numeric code is implemented in \texttt{C++}
and interfaced with \texttt {Python} into \texttt{astromodels}
\citep{Vianello:2018b}.

The overall emission is characterized by five parameters: $B$,
$\gmin$, $\gcool$, $\gmax$, and $p$. However, a strong co-linearity
exists between $B$ and $\gmin$ as their combination sets the peak of
the photon spectrum. Thus, both parameters serve as an energy scaling
which forces the setting of one of the parameters. We chose to set
$\gmin=10^5$ though the choice is arbitrary and does not affect our
results. It is therefore important to note that all parameters are
determined relatively, that is, the values of $\gcool$ and $\gmax$ are
determined as ratios to $\gmin$. Similarly, the value of $B$ is only
meaningful when determining the characteristic energies of $\gcool$
and $\gmax$ or $\hnuc$ and $\hnumax$ respectively. In other words,
with our parameterization the spectra are scale free. The degeneracies
can be eliminated by specifying temporal and radial properties of the
GRB outflow which we have neglected in this analysis.

Ideally, we would fit for the full set of parameters in the
model. However, the already high-dimensionality of the model does not
allow us to fit for the cooling regime of the model simultaneously
with the polarization due to computational time
constraints. Therefore, we first fit the spectral data alone to
determine the amount of cooling present in the data. All spectra were
found in the slow-cooling regime
\citep{Sari:1998aa,Beniamini:2013aa}. Thus, we fixed the ratio of
$\gcool$ to $\gmin$ during the full fits to the slow-cooling
regime. Tests revealed that the cooling had no impact on the recovered
polarization parameters. Additionally, the lack of high-energy data
(via the \textit{Fermi}-LAT) forces us to fix $\gmax$ such that the
synchrotron cutoff is above the spectral window. We obtain three
parameter fit for the spectrum: $B$, $p$ and the arbitrary spectral
normalization ($K$). $B$ and $K$ are given uninformative scale priors
and $p$ a weakly-informative, Gaussian prior centered around
$p=3.5$. The effective area constants applied to the POLAR response
are given truncated Gaussian priors centered at unity with a width of
10\% to reflect our prior belief that the instruments are
well-calibrated to one another\footnote{This belief will be
    conditioned on the data and thus can be modified.}.

\section{Results}
\label{sec:results}
In the following two sections, we present the results from the
combined polarization and spectral analysis separately. Corner plots
of the important (non-nuisance) parameter marginal distributions are
displayed in Appendix \ref{sec:params}.

\subsection{Polarization}
\begin{figure*}
  \centering
  \includegraphics{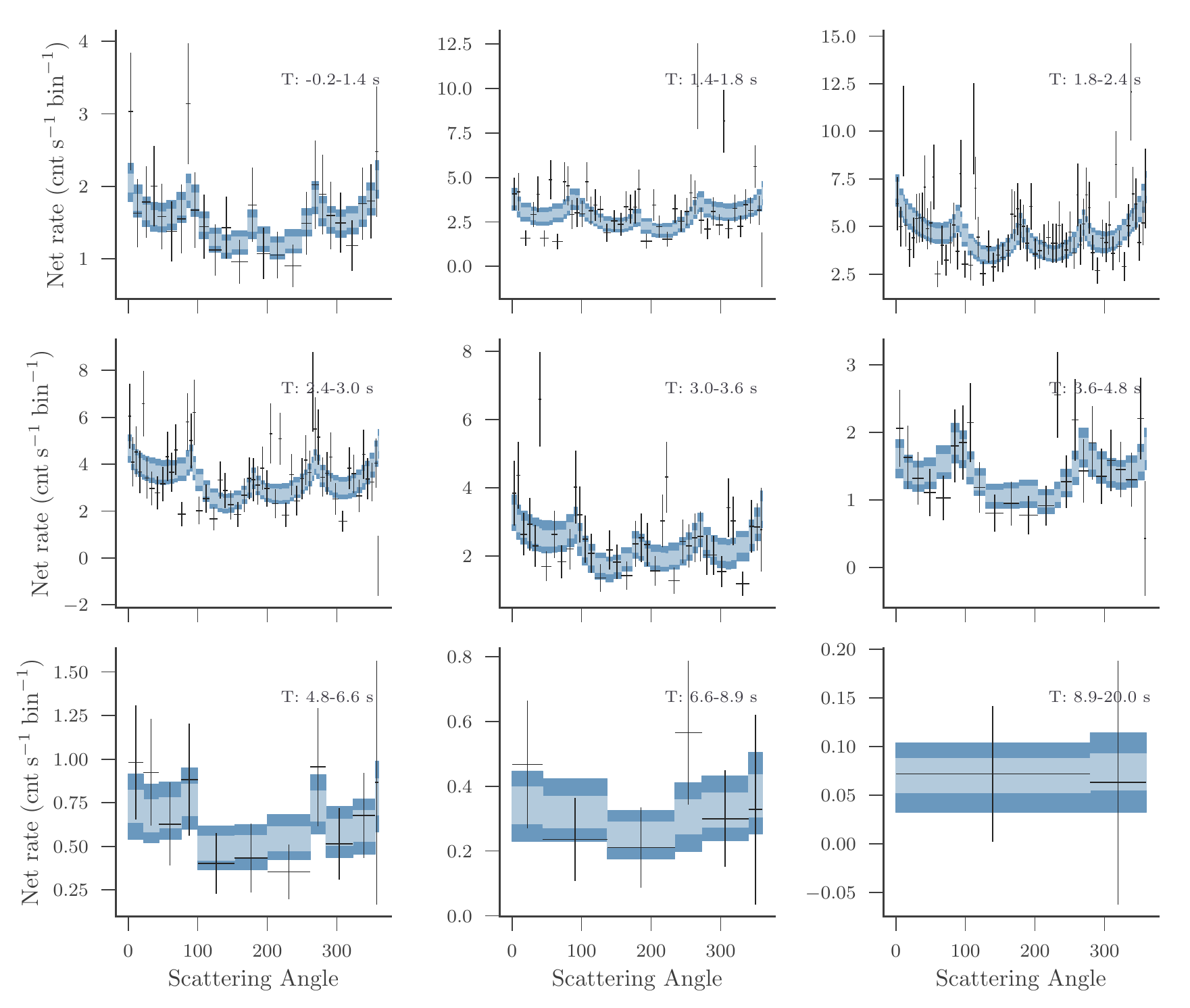}
  \caption{Net SA data for each fitted time interval
    in our analysis. Superimposed are the posterior model predictions
    from the fits. The data have been rebinned for visual clarity. The
    SA presented here is measured within an arbitrary
    local coordinate system of POLAR.}
  \label{fig:polarization_counts}
\end{figure*}

\begin{figure*}
  \centering
  \includegraphics{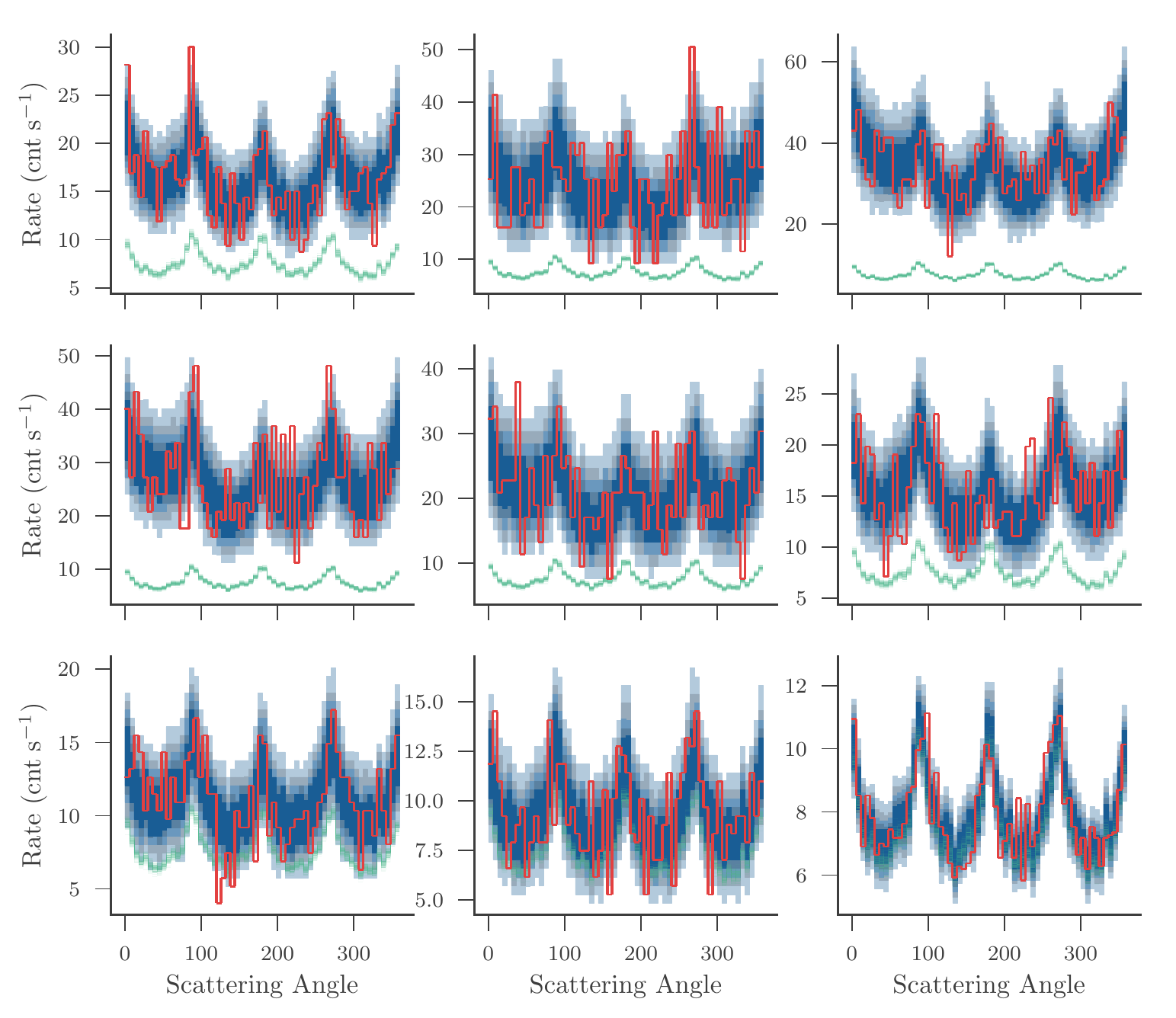}
  \caption{Posterior predictive checks for the total polarization
    count rate data. The dark to light blue shaded regions indicate
    the 50$^{\mathrm{th}}$, 60$^{\mathrm{th}}$, 70$^{\mathrm{th}}$,
    80$^{\mathrm{th}}$, and 90$^{\mathrm{th}}$ percentiles of the
    replicated data respectively of the replicated data. The observed
    data are displayed in red. The estimated background count rates
    are displayed in green}
  \label{fig:ppc}
\end{figure*}

The POLAR polarization data are well described by our modeling of the
POLAR instrument. The scattering angle data show good agreement between
the data and the model as demonstrated in Figure
\ref{fig:polarization_counts}. In order to validate the model's
ability to generate the data, we performed posterior predictive
checks (PPCs) \citep{Betancourt:2018aa} of the polarization data for
all time intervals. For a subset of posterior samples chosen with
appropriate posterior probability, latent polarization and spectral
models were generated and subsequent data quantities where sampled
from the likelihood. The model was able to sufficiently generate
replicated data similar to the observed (see Figure \ref{fig:ppc}) in
most cases. It is likely that minor deficiencies still exist in the
instrumental responses.

\begin{figure*}
  \centering
 \includegraphics[]{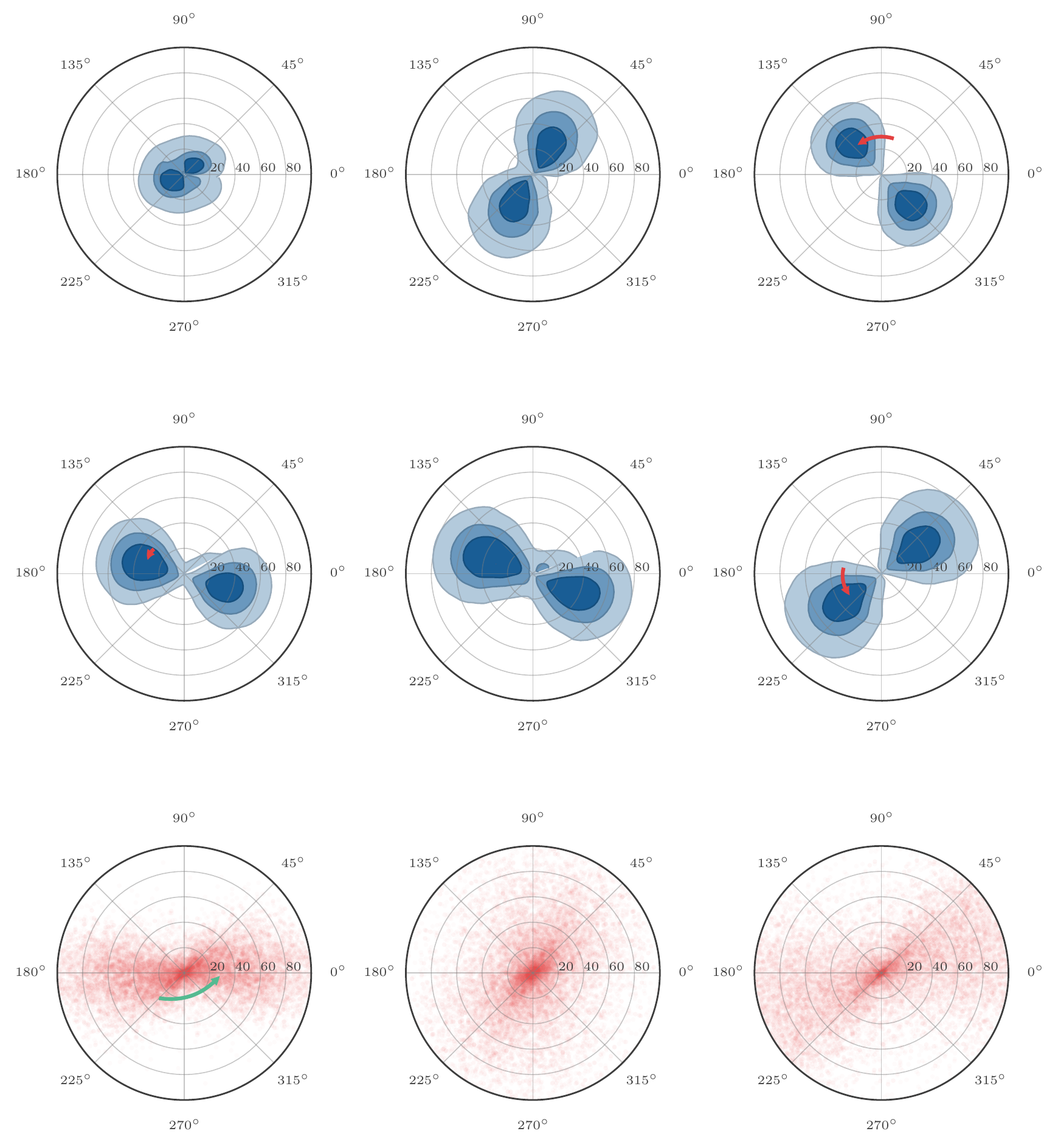}
 \caption{Posterior polarization results. The radial coordinate
   represents polarization degree and the angular coordinate the
   polarization angle. The polarization angle here is transformed to
   equatorial coordinates. The contours are for the
   30$^{\mathrm{th}}$, 60$^{\mathrm{th}}$, and 90$^{\mathrm{th}}$
   percentiles of the credible regions. The plots are reflected about
   the periodic boundary of 180$^{\circ}$ for visual clarity. For the
   last three time intervals, we do not display contours and instead
   show the posterior samples as the parameters are poorly
   identified. The arrows that point from the last to the current
   position are meant as visual guides only. }
  \label{fig:polarization}
\end{figure*}

The polarization observed here is compatible with that
  presented in \citep{POLAR2018} where an unpolarized flux was
  excluded for this single pulse GRB with 99.7\% confidence. The
  analysis presented here does, however, allow us to study the time
  evolution in significantly more detail. This is because, unlike in
the study  \citep{POLAR2018}, the polarization degree is not forced to be equal
  over all the studied time intervals but is instead left as a free
  parameter, while the number of studied time bins is increased from three
  to nine. Despite this significant increase in free parameters
  constraining measurements can still be performed. We observe no
polarization at the beginning of the pulse and moderate ($\sim 30$\%)
polarization as time proceeds. Interestingly, we observe a large
change in the polarization angle with time (see Figure
\ref{fig:polarization}). Although the time intervals used in this
study are different from those used in \citep{POLAR2018}, it can be
deduced that the polarization angles found here agree with those in
\citep{POLAR2018}. The end of the pulse has a relatively weak signal
and thus poorly identified polarization parameters. The 68\% credible
regions are listed in Table \ref{tab:table1}. Clearly, the level of
polarization during the peak of the emission is probabilistically
equivalent to both moderate, low or even 0\% polarization during
several intervals whereas during the beginning of the emission the
polarization is definitely low even though the ratio of background to
total signal is high.

We stress that it is not appropriate to perform model comparison on
nested model parameters, for example, comparing between zero polarization and
greater than zero polarization. This includes the use of Bayes factors
\citep{Chattopadhyay:2017aa} which are ill-defined for improper priors
and for comparing between discrete values of a continuous parameter
\citep{Gelman:2013}. Polarization is not a detected quantity, but a
parameter. Given that we have detected the GRB, it is important to
quote the credible regions of the polarization parameter rather than
perform model comparison between discrete values.

\renewcommand{\arraystretch}{1.5}%

\begin{table*}
  \centering
\label{tab:table1}
\caption{Parameters with their 68\% credible regions. Here $\bar{p}$ is the polarization degree (in $\%$), $\phi$ the polarization angle (in deg.), the spectral parameter $h\nu_{\mathrm{inj}}$ (in arbitrary units) and $p$ the power law index (in arbitrary units). }
\begin{tabular}{ccccc}
  \hline\hline 
    Time Interval & $\bar{p}$                 & $\phi \; (\mathrm{deg})$    & $h\nu_{\mathrm{inj}}\; (\mathrm{keV})$ & $p$                     \\
    \hline
    -0.2-1.4      & $13.21^{+6.10}_{-13.20}$  & $71.86^{+80.54}_{-49.87}$   & $362.46^{+59.34}_{-53.86}$             & $3.67^{+0.39}_{-0.57}$  \\
    1.4-1.8       & $24.19^{+10.53}_{-23.25}$ & $61.12^{+29.47}_{-24.98}$   & $242.58^{+33.76}_{-31.49}$             & $3.91^{+0.42}_{-0.52}$  \\
    1.8-2.4       & $30.10^{+16.37}_{-15.50}$ & $132.12^{+15.66}_{-15.57}$  & $268.89^{+24.96}_{-24.50}$             & $4.68^{+0.54}_{-0.55}$  \\
    2.4-3.0       & $28.29^{+16.58}_{-20.44}$ & $155.09^{+15.82}_{-134.21}$ & $160.89^{+20.35}_{-17.59}$             & $3.52^{+0.25}_{-0.36}$  \\
    3.0-3.6       & $28.62^{+12.04}_{-28.61}$ & $146.19^{+22.07}_{-113.64}$ & $110.83^{+18.42}_{-15.64}$             & $3.01^{+0.24}_{-0.26}$  \\
    3.6-4.8       & $33.45^{+15.89}_{-26.39}$ & $38.89^{+21.08}_{-16.01}$   & $62.31^{+8.90}_{-7.97}$                & $2.67^{+0.10}_{-0.15}$  \\
    4.8-6.6       & $38.26^{+15.56}_{-38.04}$ & $51.14^{+117.74}_{-40.09}$  & $103.97^{+15.64}_{-14.74}$             & $4.11^{+0.45}_{-0.59}$  \\
    6.6-8.9       & $34.90^{+15.99}_{-34.86}$ & $66.94^{+66.44}_{-40.46}$   & $59.99^{+11.56}_{-10.32}$              & $3.75^{+0.38}_{- 0.46}$ \\
  8.9-20.0        & $51.53^{+38.26}_{-26.99}$ & $46.18^{+110.32}_{-30.12}$  & $54.25^{+12.28}_{-10.73}$              & $3.83^{+0.46}_{-0.60}$  \\
  \hline
\end{tabular}
%\caption{Parameter 68\% credible regions}
\end{table*}

\subsection{Spectra}

\begin{figure*}
  \centering
  \includegraphics[]{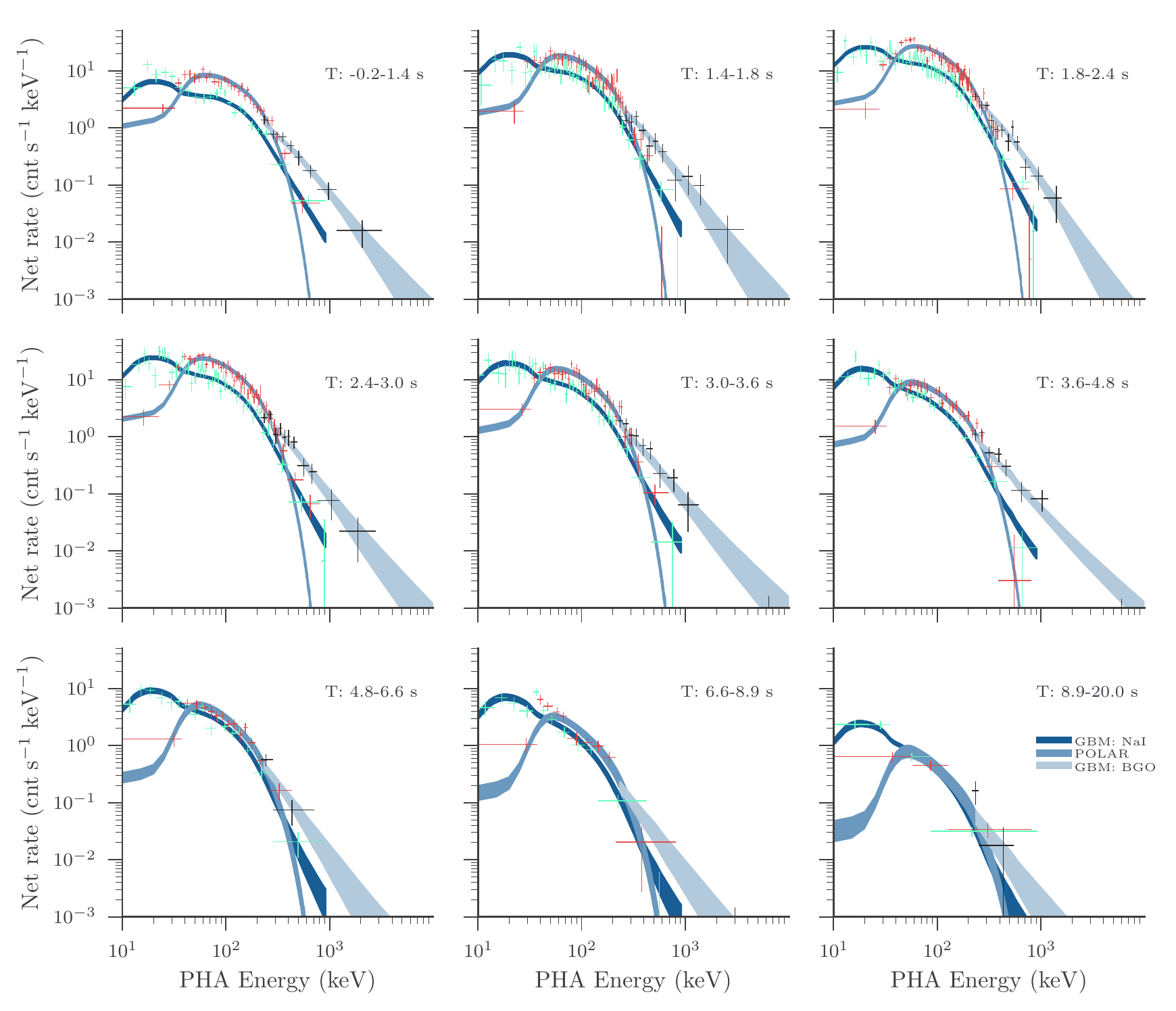}
  \caption{Count spectra of POLAR and GBM from the joint
    spectral and  polarization fits. The shaded regions indicate the
    2$\sigma$ credible regions of the fit. Data from a GBM NaI, BGO,
    and POLAR and displayed in green, black, and red respectively.}
  \label{fig:counts}
\end{figure*}

POLAR and GBM observed data both agree in overall spectral shape and relative
normalization of the observed flux. Moreover, the spectral results
demonstrate that the synchrotron spectrum is a good, predictive
description of the spectral data as displayed in
Figure~\ref{fig:counts}. This is both a confirmation that past studies
with synchrotron relying on GBM data alone are reliable, as well as the
outstanding calibration between the GBM and POLAR.

\begin{figure}[h!]
  \centering
  \includegraphics{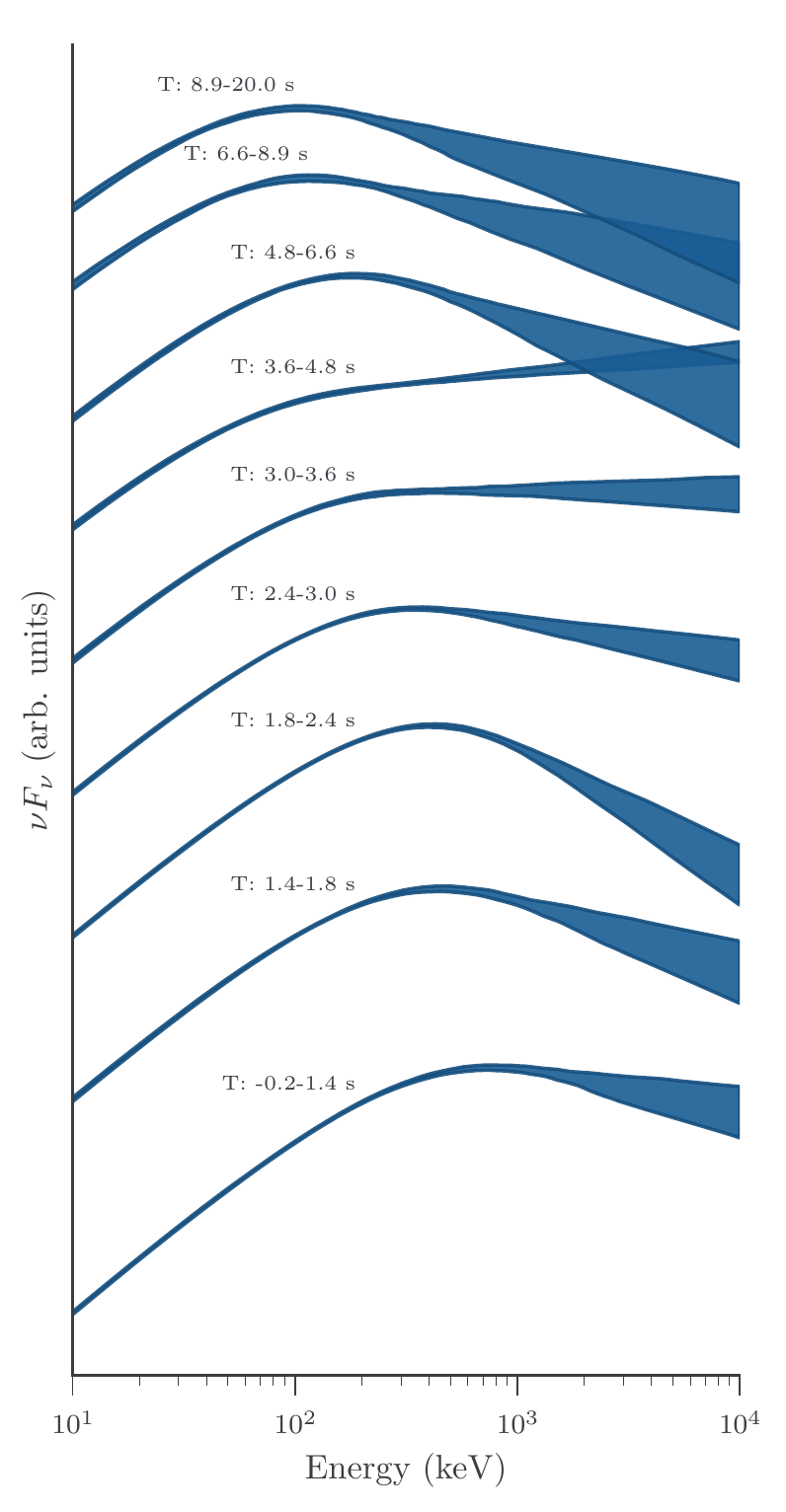}
  \caption{ $\vFv$ spectra of the synchrotron fits scaled with
    increasing time. The width of the curves represents the 1$\sigma$
    credible regions of the model.}
  \label{fig:spec}
\end{figure}

As noted above, it is not possible to disentangle the intrinsic
parameters of the synchrotron emission without further
assumptions. Therefore, we only quote the injection energy in Table
\ref{tab:table1}. The evolution of the spectrum is shown in Figure
\ref{fig:spec}. The temporal evolution of the $\vFv$ spectral peak
follows a broken power law. We find values between approximately three
and four for the electron power law
injection spectral index. These values are steeper than those of the
canonical index expected from shock acceleration \citep{Kirk:2000vr}.

It is possible that other physical spectral models also provide
acceptable, predictive, fits to the data. However, these models -- for example
subphotospheric dissipation -- have yet to demonstrate acceptable
spectral fits on a large sample of GRBs. Moreover, the numerical
schemes \citep{Peer:2005aa} required to compute the emission form
these models are more complex than that of our synchrotron modeling,
require far more computational time, and are not publicly available
for replication. Photospheric models also require special geometrical
setups to produce polarization. This makes them more predictive, and
indeed a pertinent set of models to test.

\section{Discussion}
\label{sec:discussion}
For the first time, the polarization and spectrum of GRB prompt
$\gamma$-ray emission has been fitted simultaneously. Furthermore, the
spectral data have been described with a physical synchrotron model
consistent with the spectral data of two very distinct
spectrometers. We argue that it is unlikely for the spectral and
polarization data to conspire to point toward an optically thin
synchrotron origin of the emission. However, the current predictive
power of GRB prompt emission polarization theory is not developed
enough for our measurements to definitively select synchrotron over
other emission mechanisms. Therefore, we speculatively leverage
previous spectral results that show that synchrotron emission is
dominant mechanism in single-pulse GRBs.

\citet{Burgess:2018} argue that the observation of synchrotron
emission in GRBs invalidates the standard fireball model
\citep{Eichler:2000aa}. Similar predictions were made before they were
supported by data \citep[e.g.,][]{Zhang:2009aa}. These results allude
to a magnetically dominated jet acceleration mechanism possibly
resulting in comoving emission sites or mini-jets
\citep{Barniol-Duran:2016aa,Beniamini:2018aa}. These results were
arrived at considering spectral analysis alone. The moderate
polarization degree observed in this work requires a development in
the prediction of the temporal polarization predictions of these
models in order to fully interpret their meaning.

\begin{figure}
  \centering
 \includegraphics[]{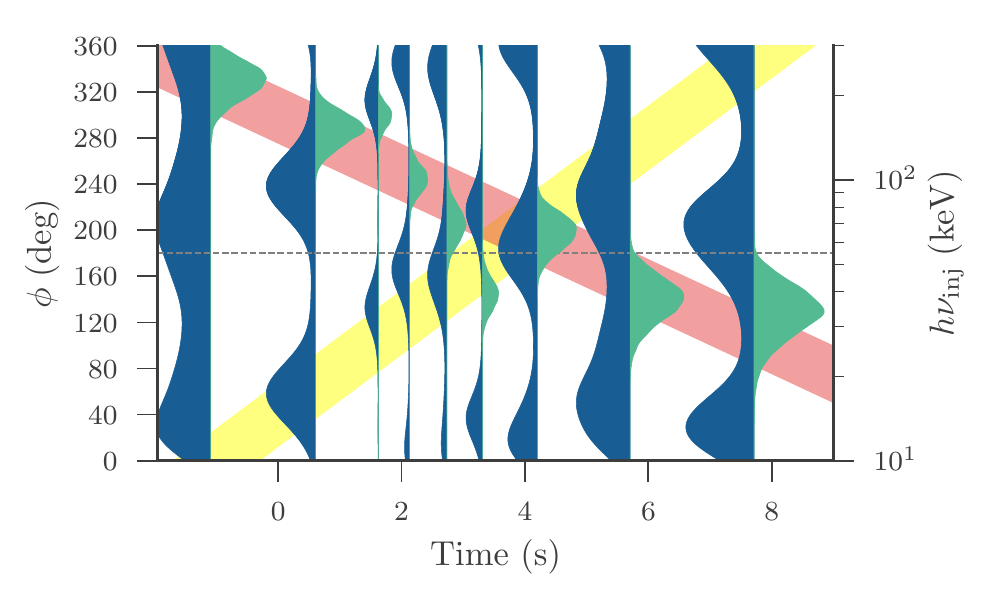}
 \caption{Temporal evolution of $h\nu+{\mathrm{inj}}$ and the
   polarization angle which has been doubled for visual
   clarity. $h\nu_{\mathrm{inj}}$ falls as the polarization angle
   increases in time. The red and yellow areas are illustrative guides for
   the evolution the parameters. No fits were performed. }
  \label{fig:polarization_fun}
\end{figure}

While our observations provide broad ranges for the observed
polarization degree, the changing polarization angle is easily
observed. Although an evolution of the polarization angle has been
reported before for multipulse GRBs using data from both the GAP and
IBIS instruments, \citep{Yonetoku:2011ef,G_tz_2009} this intrapulse
evolution has not been observed before. Figure
\ref{fig:polarization_fun} shows the way in which the both the peak of
the synchrotron spectrum and the polarization angle grossly track each
other in time. Detailed model predictions for the evolution of the
polarization angle during the GRB are not available. We are therefore
not able to interpret the change in angle and encourage the community
to develop detailed predictions which can be fitted to our data in the
future. With more predictive models, appropriate informative priors
can be adopted. Moreover, spectral parameters can be formulated in
terms of polarization parameters making the models stricter and the
data more useful. Thus, we are hopeful that models are
  developed in the near future.

The combination of POLAR and GBM observations of GRBs enables
energy-dependent polarization measurements and is a project currently
under development. These measurements will allow us to decipher if
polarization increases around the peak of the photon spectrum which
would be a signature of synchrotron emission, or if the polarization
is higher at low energies as predicted by \citet{Lundman:2018aa}. We
encourage researchers to carry out further multimessenger studies and
missions to answer these questions.

\section{Software availability}
\label{sec:software}
The analysis software utilized in this study are primarily
\texttt{3ML} and \texttt{astromodels}. We have designed a generic,
preliminary, polarization likelihood for similar X-ray polarization
instruments both within
\texttt{3ML}\footnote{\url{https://github.com/giacomov/3ML/tree/master/threeML/utils/polarization}}
and
\texttt{astromodels}\footnote{\url{https://github.com/giacomov/astromodels/blob/master/astromodels/core/polarization.py}}. Additionally,
the POLAR
pipeline\footnote{\url{https://github.com/grburgess/polarpy}} we have
developed is fully designed to be easily modified for other
instruments with polarimetric data. We note that these software
distributions are preliminary, and we encourage the community to
participate in their development.

\begin{acknowledgements}
  JMB acknowledges support from the Alexander von Humboldt
  Foundation. MK acknowledges support by the Swiss National Science
  Foundation and the European Cooperation in Science and
  Technology. The authors are grateful to the \textit{Fermi}-GBM team
  and HEASARC for public access to \textit{Fermi} data products. We
  thank Damien B\'{e}gu\'{e}, Dimitrios Giannois, Thomas Siegert and
  Ramandeep Gill for fruitful discussions.
%We thank Damien Begue and Dimitrios Giannois for enlightening discussions.
\end{acknowledgements}

\bibliographystyle{aa}
\bibliography{bib}

\appendix

\section{The POLAR polarization response}
\label{sec:polar_response}

The POLAR instrument is described in full detail in
  \citep{Produit2018}. The instrument design is such that issues found
  in previously reported polarization measurements are mitigated, for
  example fast electronics allows to record events within a
  $50\,\mathrm{ns}$ coincidence window, thereby removing chance
  coincidence induced events which can induce fake polarization. The
POLAR response was modeled using the POLAR simulation software
presented in \citet{Kole2017} which was previously used for the
analysis presented in \citep{POLAR2018}. The spectral and polarization
response are produced using the same simulation set.  Different event
selections were applied to produce the spectral and polarization
response; whereas all clean photon-like triggers, as defined in
\citet{LI2018}, were used for the spectral response, additional cuts
are applied in the event selection for the polarization response. This
causes the count rate to be higher in the spectral light curve than in
the polarization light curve as seen in Figure \ref{fig:lightcurve}.  The
selection criteria for polarization events are equal to those
previously used in \citet{POLAR2018}. In this event selection only
triggers containing at least two energy depositions in non-neighboring
bars are selected.

Simulations were performed for a grid of polarization parameters with
steps of three degrees in $\phi$ while for $\pfrac$ only 0 and 100\% were
simulated. All additional values of $\pfrac$ on the grid can be
produced by combining these results. Such a grid in polarization space
was produced for photon energies in the range of 30 to 850 keV in
steps of 5 keV, thereby producing a 3D grid of the instrument
response. The final result of each simulation is a binned modulation
curve, with a total number of 360 bins, normalized to the effective
area of POLAR for the specific photon energy. We note here
that the effective area is found to be independent of the
polarization, as could be naively expected. Therefore, the
polarization sensitivity is proportional to the source counts, and
thus highest in the $\sim$50-150 keV range.

Uncertainties in the simulated response are taken into account by
adding an additional uncertainty to each bin in addition to that
coming from the statistical uncertainty. As presented in
\citep{LI2018} the main uncertainty in the response stems from
uncertainties of the gain calibration. The propagation of this
uncertainty to the polarization response was studied here and is found
to result in a typical relative uncertainty of $2\%$ for each bin in
the polarization response. All other uncertainties, such as those from
other calibration parameters or uncertainties in the mass model of
both POLAR and the surrounding materials, are found to be negligible,
as previously presented in \citep{LI2018, POLAR2018}. The systematic
uncertainties in the polarization stemming from spectral uncertainties
are naturally included by fitting for the spectrum and the
polarization at the same time. Finally, unlike in the results
presented in \citep{POLAR2018} the location induced uncertainty is
negligible in this analysis due to the highly precise location
acquired using the BALROG.

\section{Polarization assessment}
\label{sec:polar_assessment}
We wish to validate our analysis method via simulations to verify that
under the assumption of the true model our inferences are
identifiable. Therefore, we created simulations of both spectra and
polarization for sets of $(\phi, \pfrac)$ and an assumed power law
spectrum and fit them with the same likelihood used for real data. To
avoid pathologies that can be introduced with energy dispersion, we
assumed an x-ray detector with an identity response and simulate a
simple power law photon spectrum. Both the POLAR response and a
simulated background are included as we are mainly concerned with
validating our polarization inferences.The background simulations were
performed by sampling events from real in-orbit data recorded by POLAR
in a period both before and after the GRB 170114A. It should be noted
here that the SA distribution of the POLAR background was found to be
very stable \citep{POLAR2018}.

\begin{figure}
  \centering \includegraphics{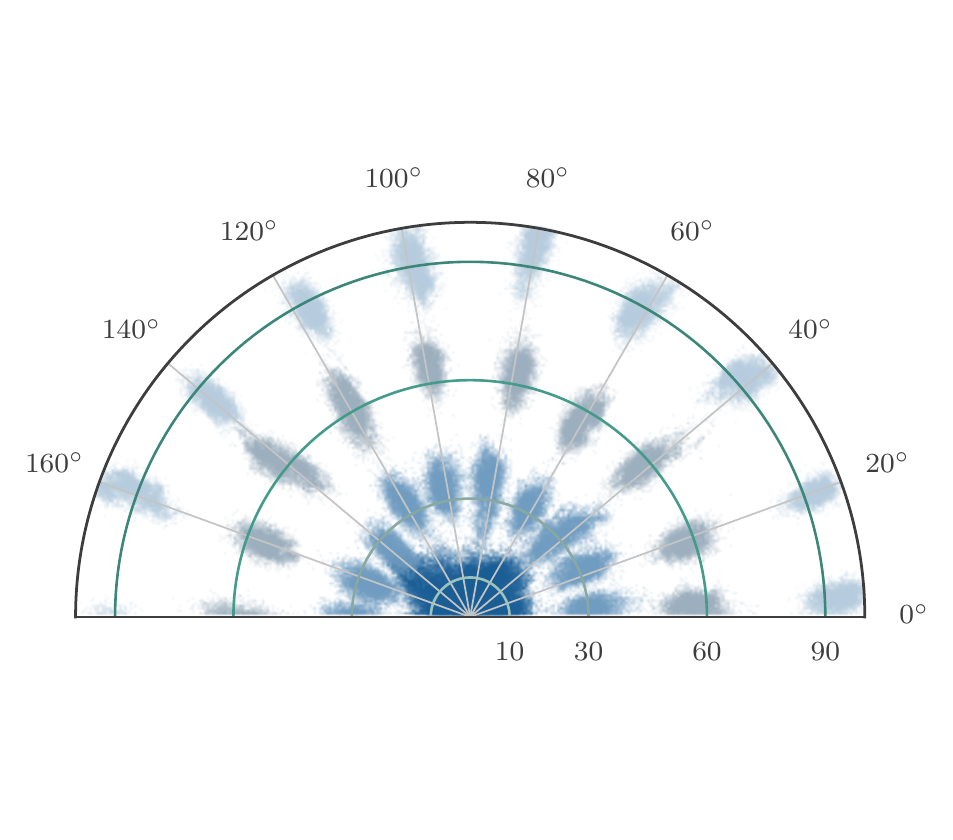}
  \caption{Posterior samples from fits to simulated polarization and
    spectral data simulated with a high signal-to-noise ratio. The
    true simulated polarization degrees and lines are demonstrated
    with green lines and gray rays respectively. The posterior samples
    are colored from dark blue to light blue with increasing simulated
    polarization degree. Thus, we demonstrate that our posteriors
    encapsulate the simulated values directly without resorting
    to the statistical approximations of past works.  }
  \label{fig:highsnsim}
\end{figure}

We simulated a nested grid of
$\phi \in \left\{0^{\circ}, 160^{\circ} \right\}$ in steps of
$20^{\circ}$ and $\pfrac \in \left\{0, 90 \right\}$ in steps of ten. We
simulated at both low and high signal-to-noise ratio levels. The partial
results from the high signal-to-noise ratio simulations are shown in Figure
\ref{fig:highsnsim}. Simulations at lower signal-to-noise ratios provide
similar results but with broader parameter credible regions. From our
simulations, we are satisfied that our construction of the likelihood
provides valid inferences.

\begin{figure*}
  \centering
  \includegraphics{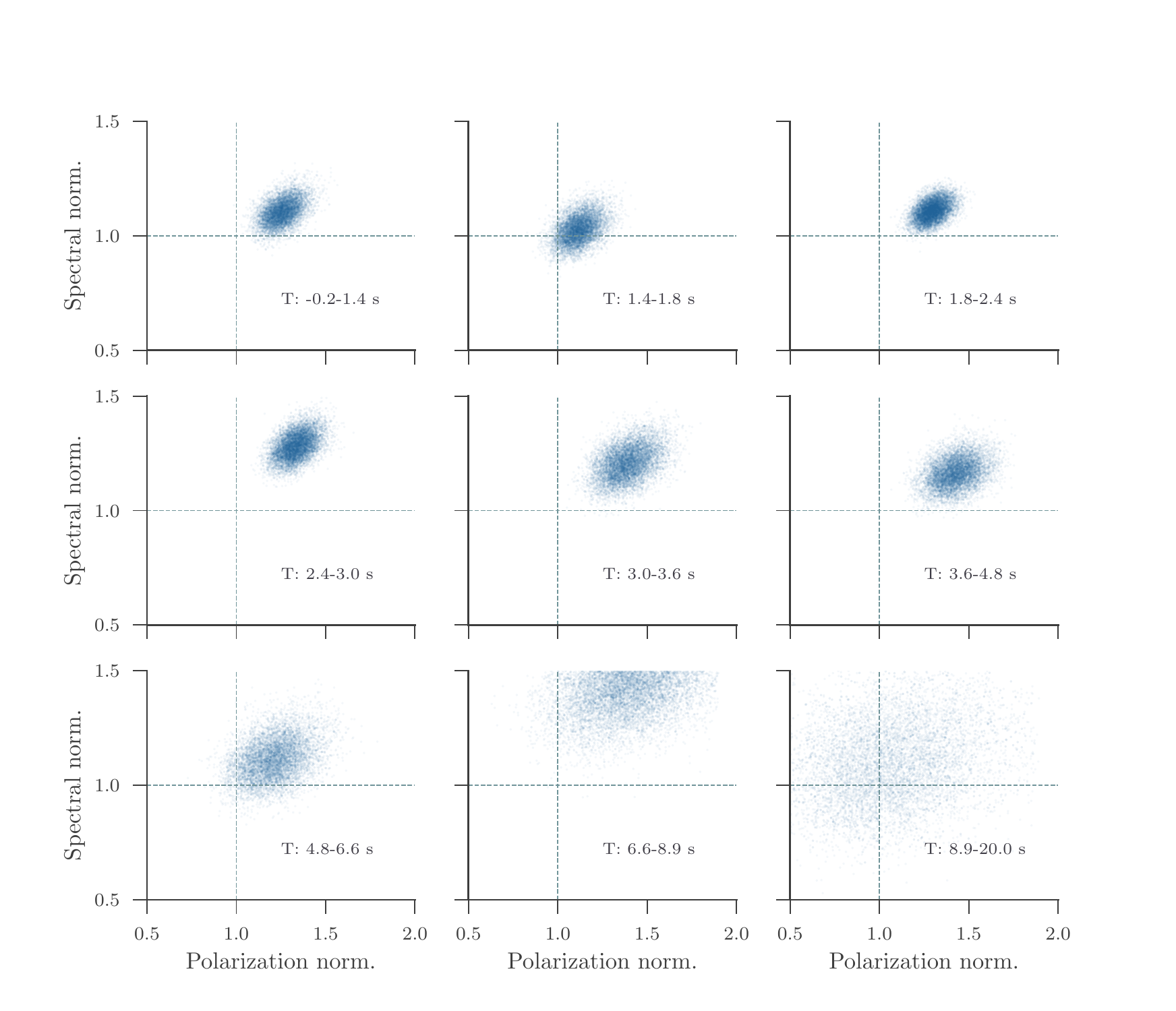}
  \caption{Spectral and polarization normalization posteriors for
    the POLAR data. There is an evident $\sim20-30$\% different in the
    relative flux between GBM and POLAR. }
  \label{fig:cal}
\end{figure*}

In Figure \ref{fig:cal} we display the posterior samples of the POLAR
polarization and spectral normalization constants. The values obtained
are not dissimilar from the values typically found between GBM
detectors when those parameters are allowed to vary in the fits
\citep{Yu:2016aa}.

\section{The likelihood}
\label{sec:likelihood}
Here we specify the full likelihood utilized in the analysis. We first
defined basic distributions, that is, the Gaussian and Poisson
distributions respectively:

\begin{eqnarray}
  \label{eq:12}
  \cond{\mathcal{N}}{x}{\mu,\sigma} & =& \frac{1}{\sqrt{2\pi \sigma}} \exp\left( -\frac{1}{2} \left( \frac{x-\mu}{\sigma} \right)^2\right)\\
  \cond{\mathcal{P}}{n}{\lambda}& = &\frac{\lambda^n e^{-\lambda}}{n!} \text{.}
\end{eqnarray}
\noindent
The data from POLAR and GBM are inherently Poisson-distributed as they
are counting experiments. The total count data in the $i^{\rm th}$
detector channel ($N_i$) are a mixture of latent source($s_i$) and
background ($b_i$) events. The transient nature of GRBs allowed us to
naively separate the source observations into temporally on- and
off-source regions. The background could then be modeled temporally in
each detector channel as a polynomial resulting in estimate of the
background counts ($B_i$) with and associated error
($\sigma_{B_i}$). Thus our data for each detector channel (both PHA
and SA channels) are the total counts $N_i$, $B_i$, and
$\sigma_{B_i}$. It is immediately obvious that we cannot simply
subtract the background counts from the data as we (i) cannot know
which counts are background and (ii) the background process has
statistical properties. Thus, we must model the joint probability of
the total and background process as
\begin{equation}
  \label{eq:13}
  \cond{\mathcal{PG}}{N_i}{s_i, b_i, B_i, \sigma_{B_i}} = \cond{\mathcal{P}}{N_i}{s_i+b_i} \cond{\mathcal{N}}{B_i}{b_i,\sigma_{B_i}} \text{.}
\end{equation}
\noindent
In our situation we did not have a spectral or polarization model for
the background process. Thus, we adopt the common procedure of
maximizing the probability with respect to $b_i$ a priori leading to
the profile likelihood referred to as PGSTAT\footnote{See \url{https://heasarc.gsfc.nasa.gov/xanadu/xspec/manual/XSappendixStatistics.html}
  or \url{https://giacomov.github.io/Bias-in-profile-poisson-likelihood/}
  for detailed discussion.}.

Thus, with $j$ datasets, that is, spectral or polarization detector's data, the total likelihood for our observations is
\begin{equation}
  \label{eq:14}
  \mathcal{L}= \prod_{j=1}^{N_{\mathrm{det}}}\prod_{i=1}^{N^{j}_{\mathrm{chan}}} \cond{\mathcal{PG}}{N^j_i}{s_i, B^j_i, \sigma_{B^j_i}} \text{.}
\end{equation}

Previous $\gamma$-ray polarization estimates have been achieved via
background-subtracted data with the assumption of a Gaussian
likelihood. This is improper and can lead to systematically biased
results. There have also been attempts to transfer the statistical
techniques used in optical polarimetry
\citep{Vaillancourt:2006aa,Quinn:2012aa} to $\gamma$-rays. These
techniques are invalid for measurements that infer latent
polarization via a secondary measurement such as the Compton
scattering angle of a photon. Moreover, these techniques assume that
none of the inherent difficulties of $\gamma$-ray photon measurement
are present, namely, low counts and dispersion both in energy and
scattering angle. We have dealt with this first issue via the proper
Poisson-based likelihood. The second issue via our modeling of the
responses of our instruments directly in the inference process.

\section{Parameter corner plots}
\label{sec:params}
\begin{figure*}
  \centering
  \includegraphics[]{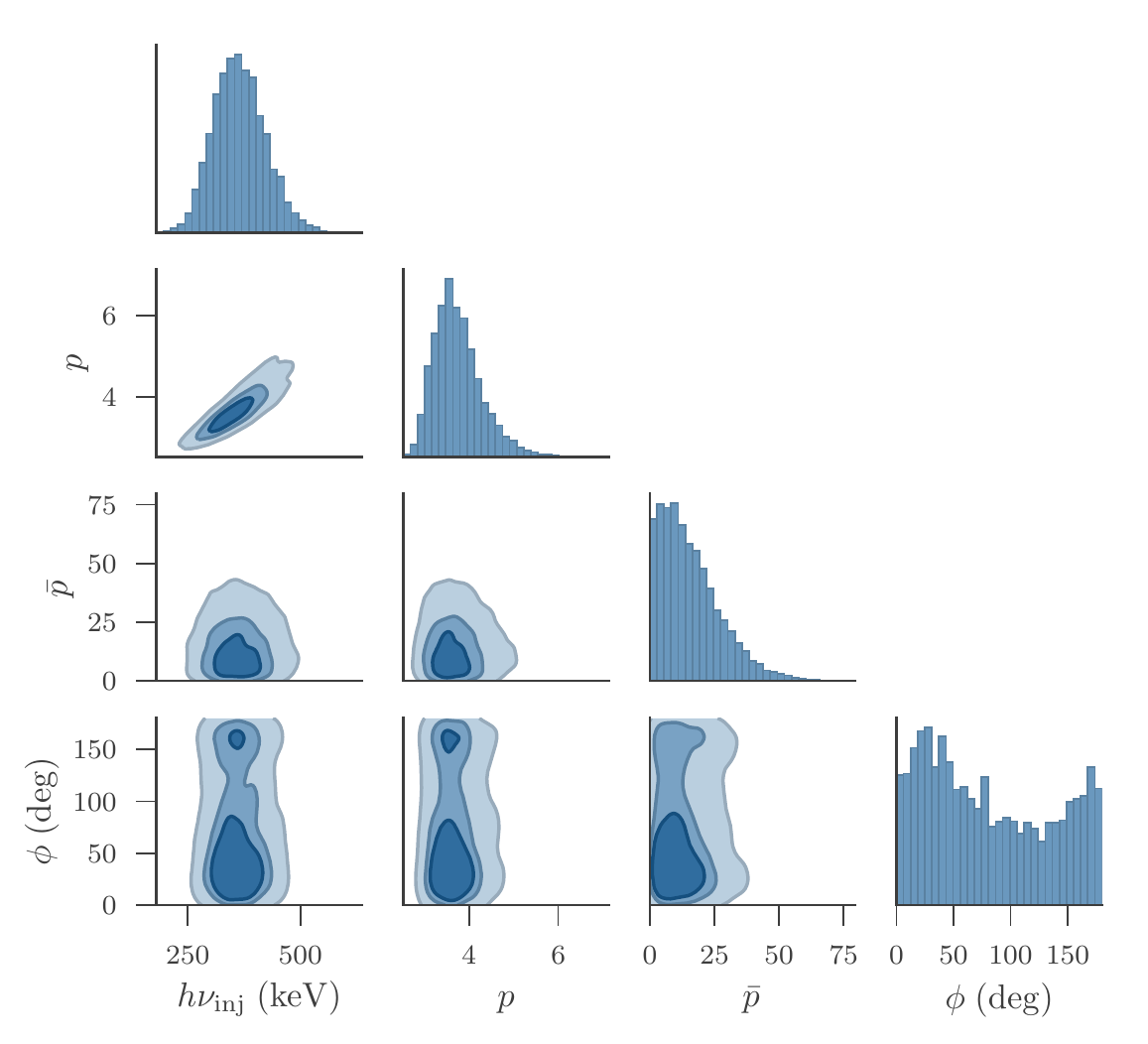}
  \caption{Parameter marginal distributions for time interval T: -0.2-1.4 s. }
  \label{fig:corner1}
\end{figure*}

\begin{figure*}
  \centering
  \includegraphics[]{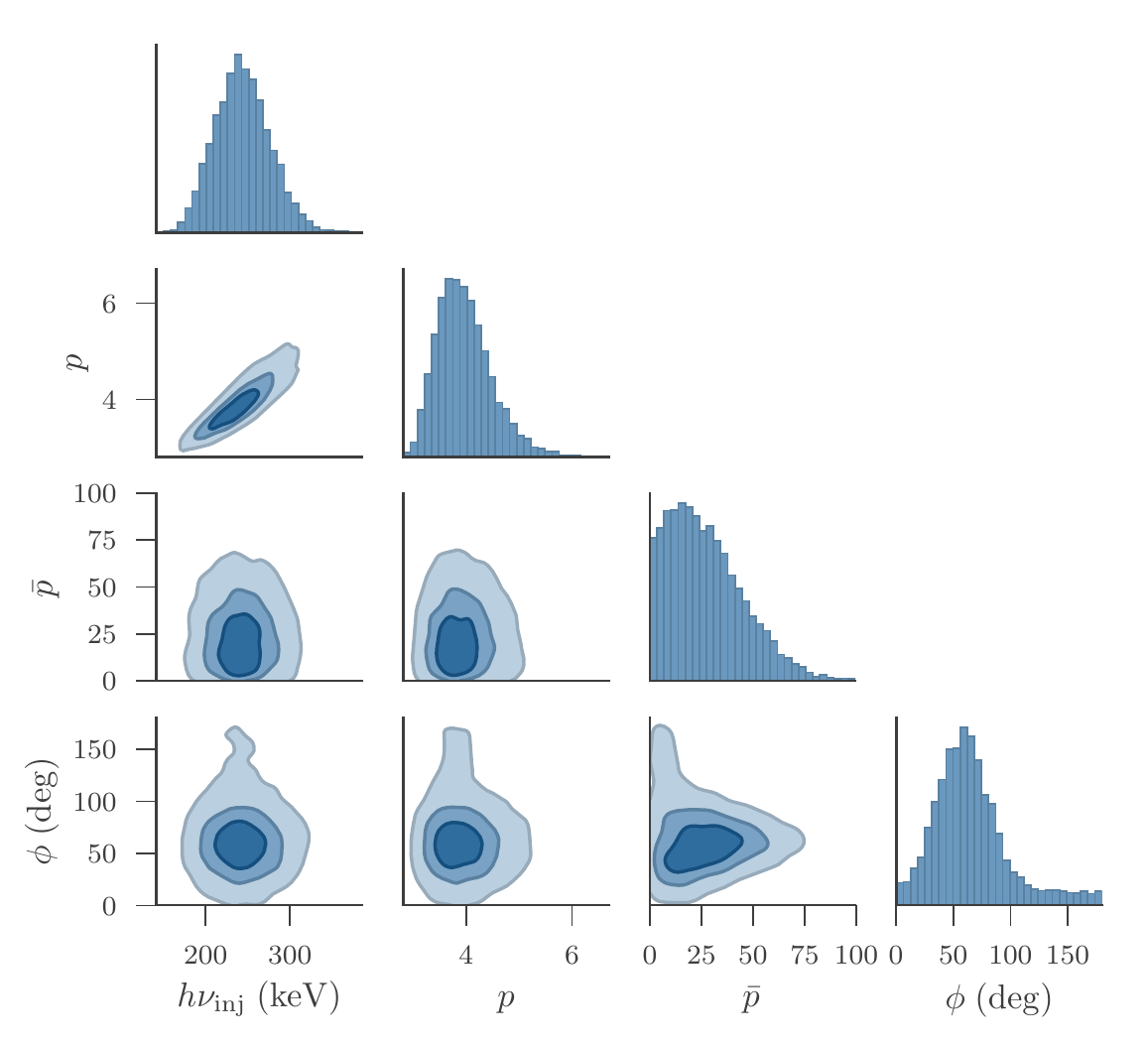}
  \caption{Parameter marginal distributions for time interval T: 1.4-1.8 s. }
  \label{fig:corner2}
\end{figure*}

\begin{figure*}
  \centering
  \includegraphics[]{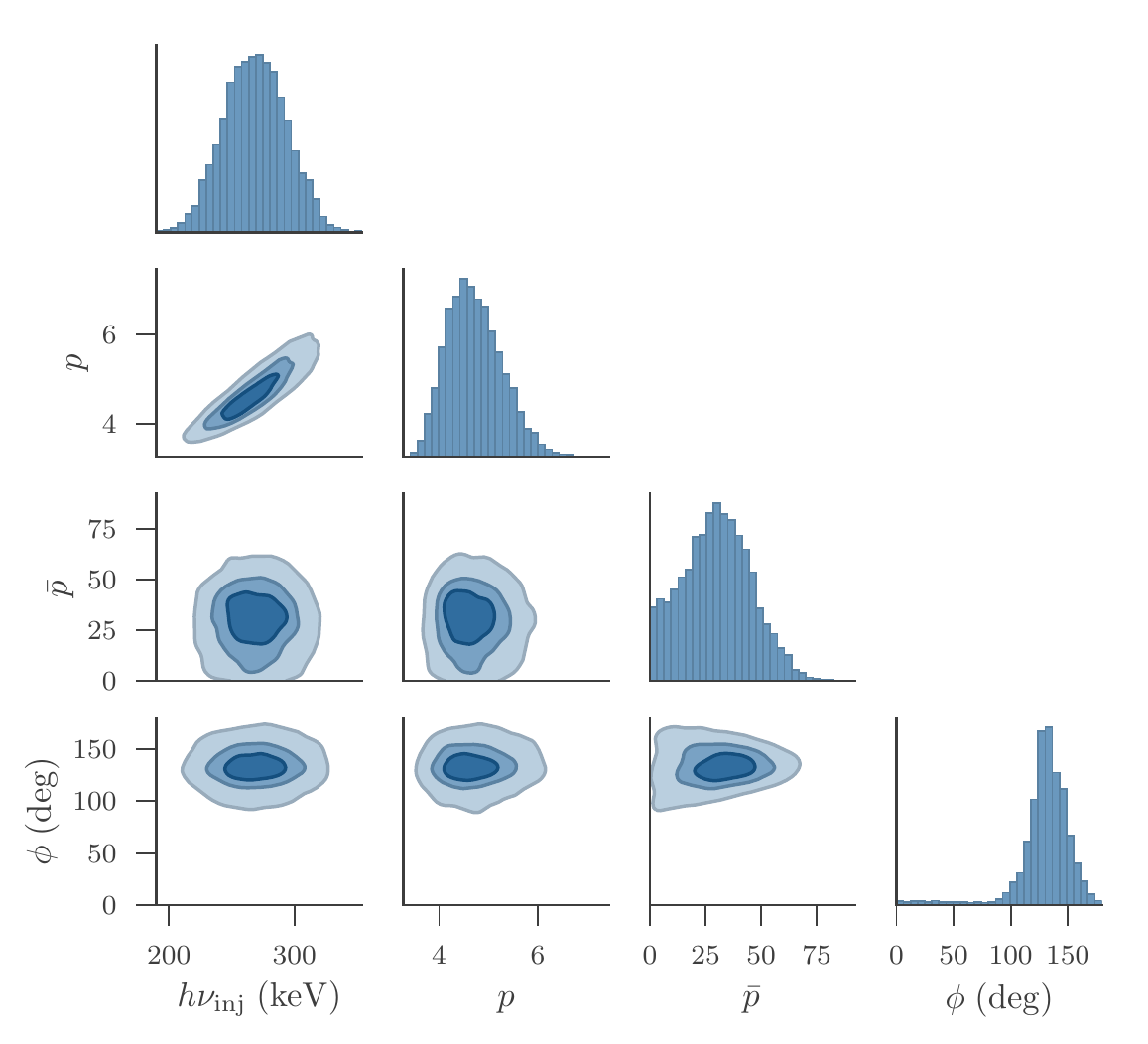}
  \caption{Parameter marginal distributions for time interval T: 1.8-2.4 s. }
  \label{fig:corner3}
\end{figure*}

\begin{figure*}
  \centering
  \includegraphics[]{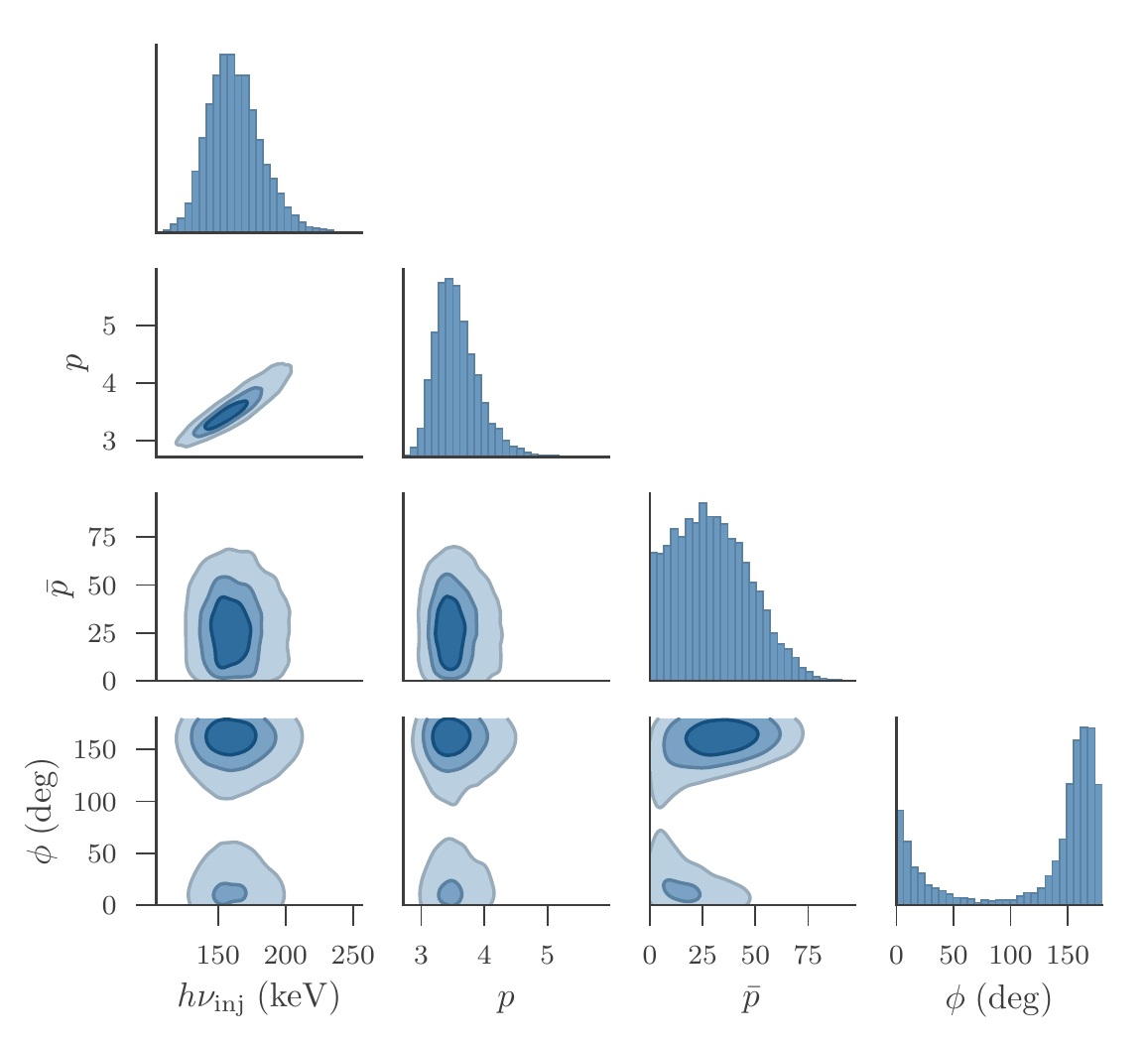}
  \caption{Parameter marginal distributions for time interval T: 2.4-3.0 s. }
  \label{fig:corner4}
\end{figure*}

\begin{figure*}
  \centering
  \includegraphics[]{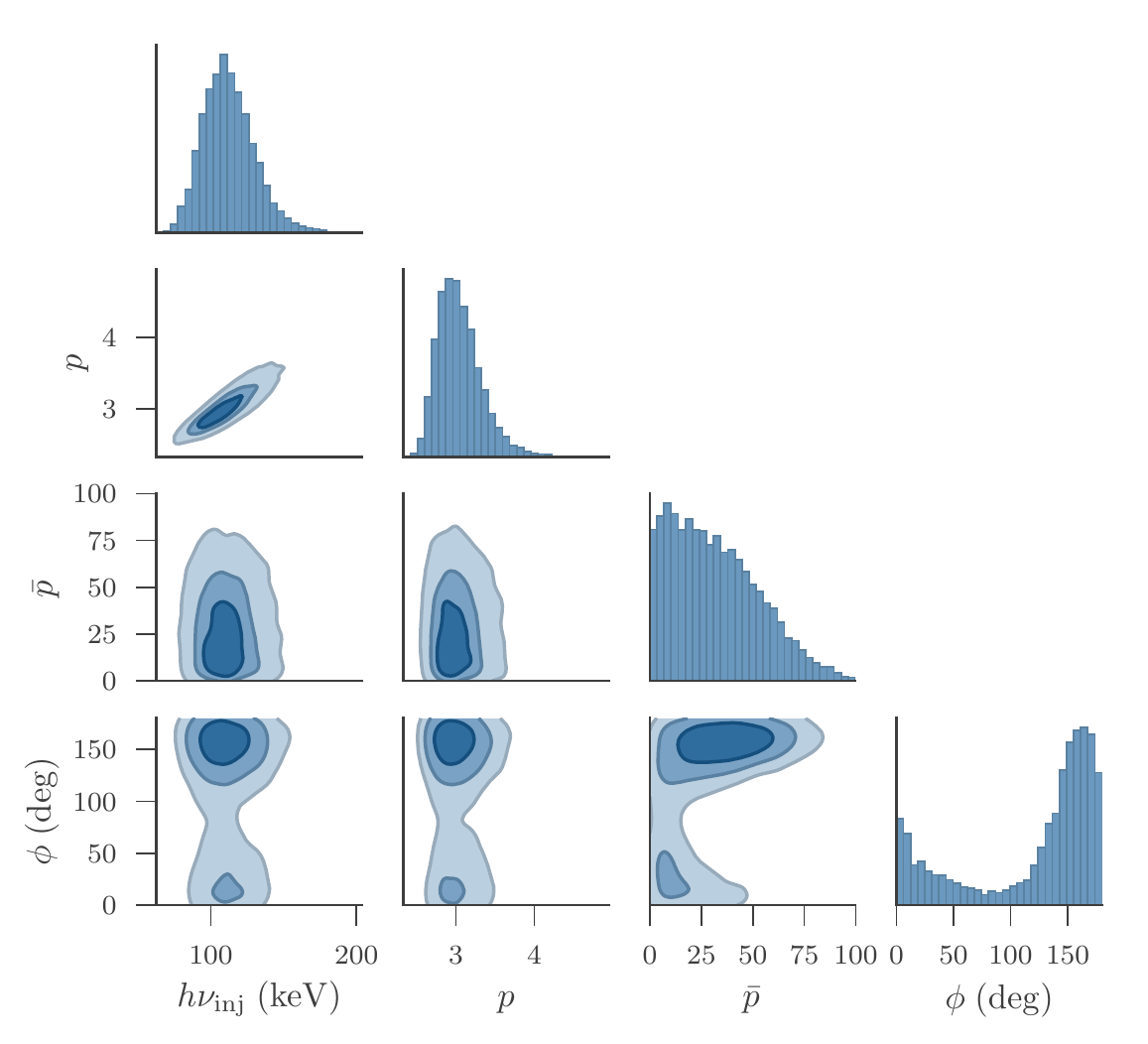}
  \caption{Parameter marginal distributions for time interval T: 3.0-3.6 s. }
  \label{fig:corner5}
\end{figure*}

\begin{figure*}
  \centering
  \includegraphics[]{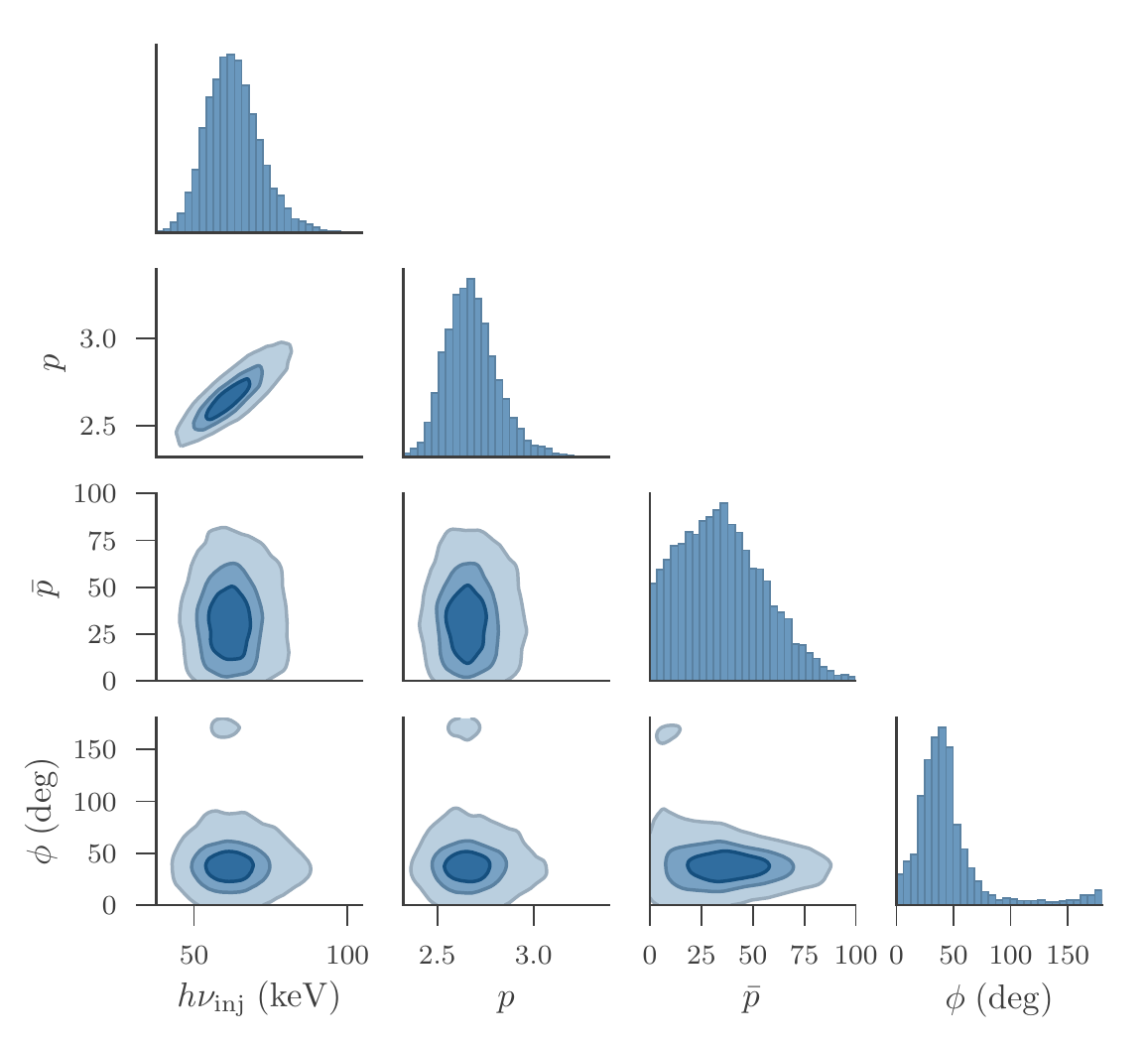}
  \caption{Parameter marginal distributions for time interval T: 3.6-4.8 s. }
  \label{fig:corner6}
\end{figure*}

\begin{figure*}
  \centering
  \includegraphics[]{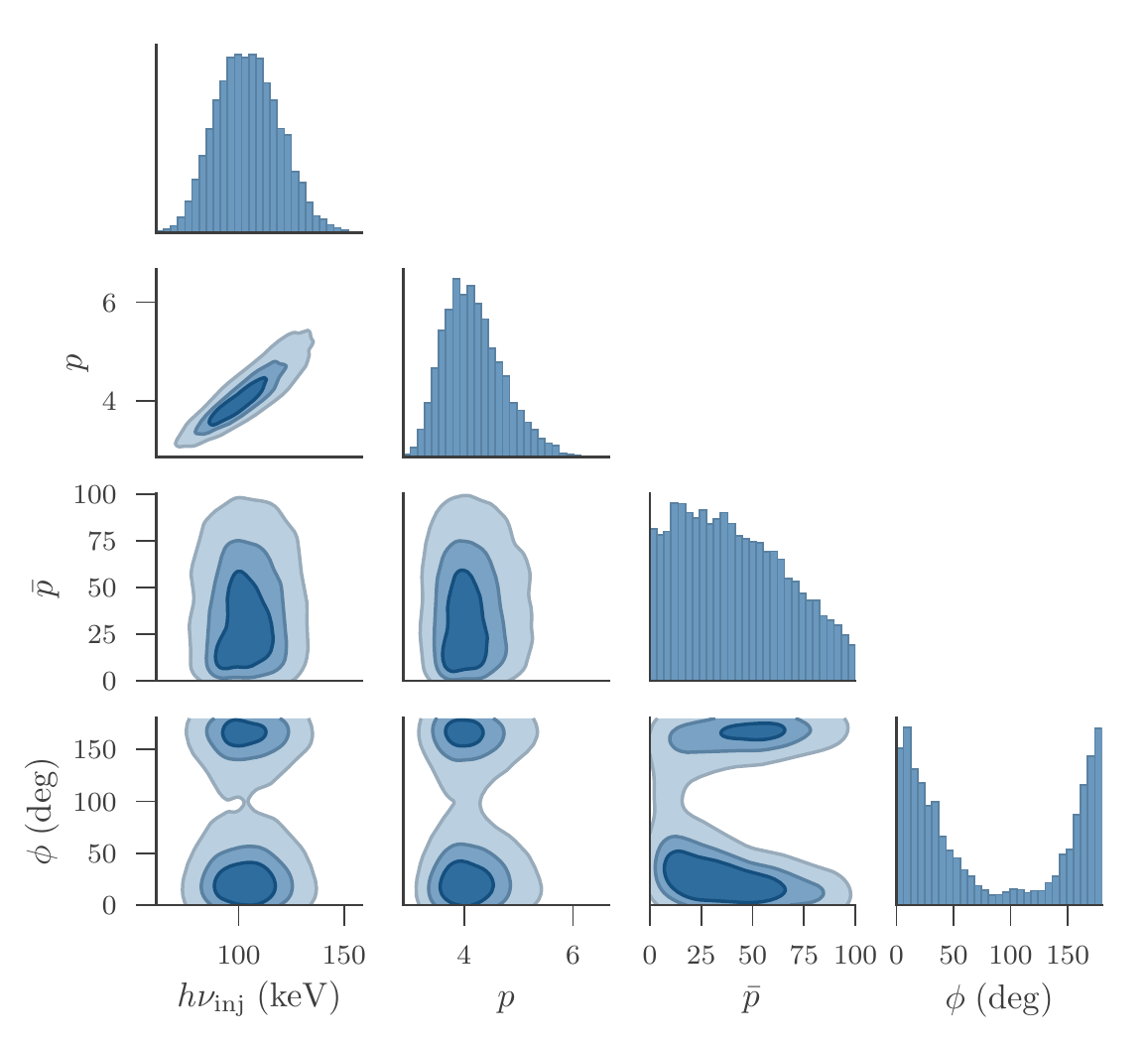}
  \caption{Parameter marginal distributions for time interval T: 4.8-6.6 s. }
  \label{fig:corner7}
\end{figure*}

\begin{figure*}
  \centering
  \includegraphics[]{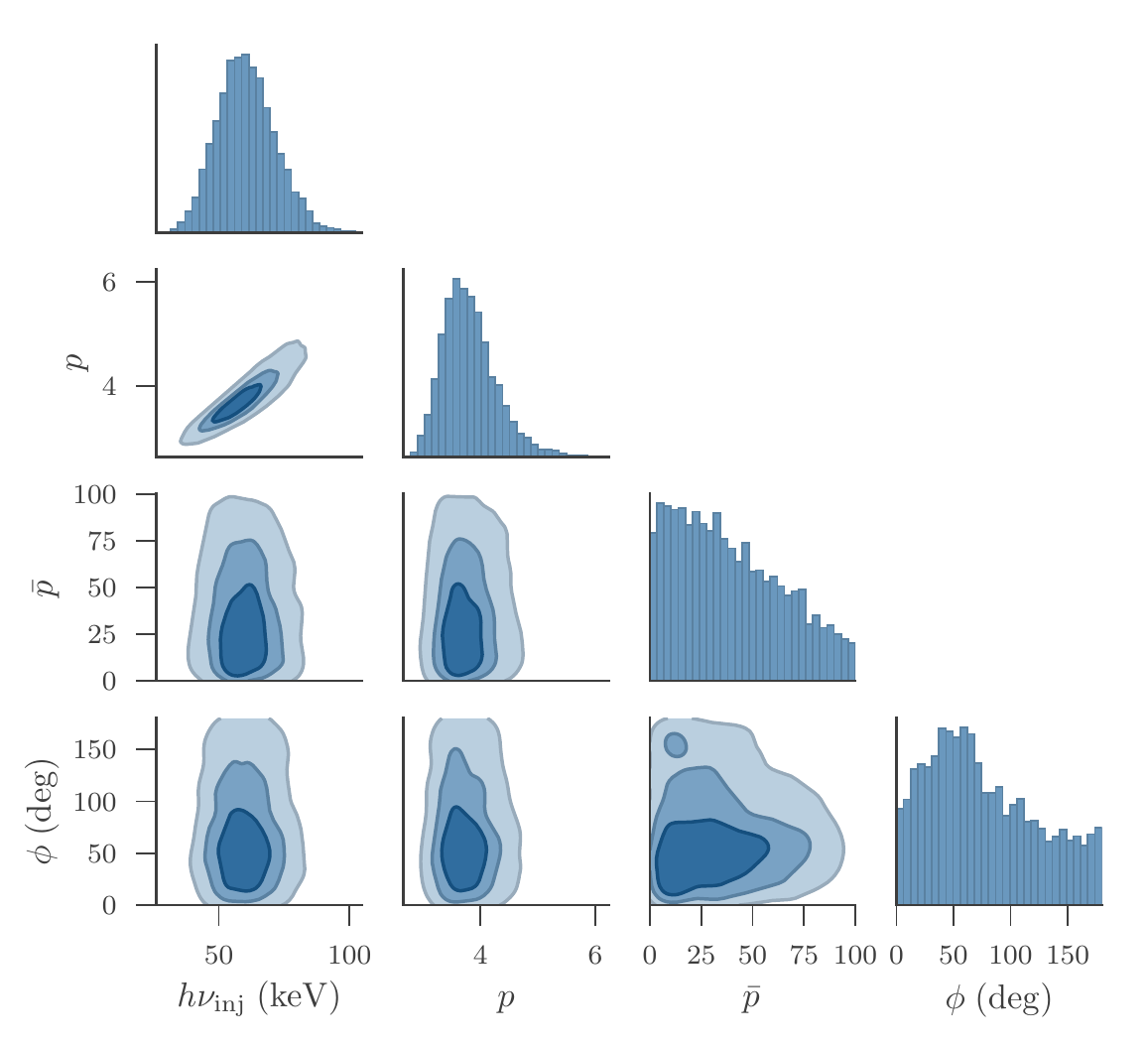}
  \caption{Parameter marginal distributions for time interval T: 6.6-8.9 s.}
  \label{fig:corner8}
\end{figure*}

\begin{figure*}
  \centering
  \includegraphics[]{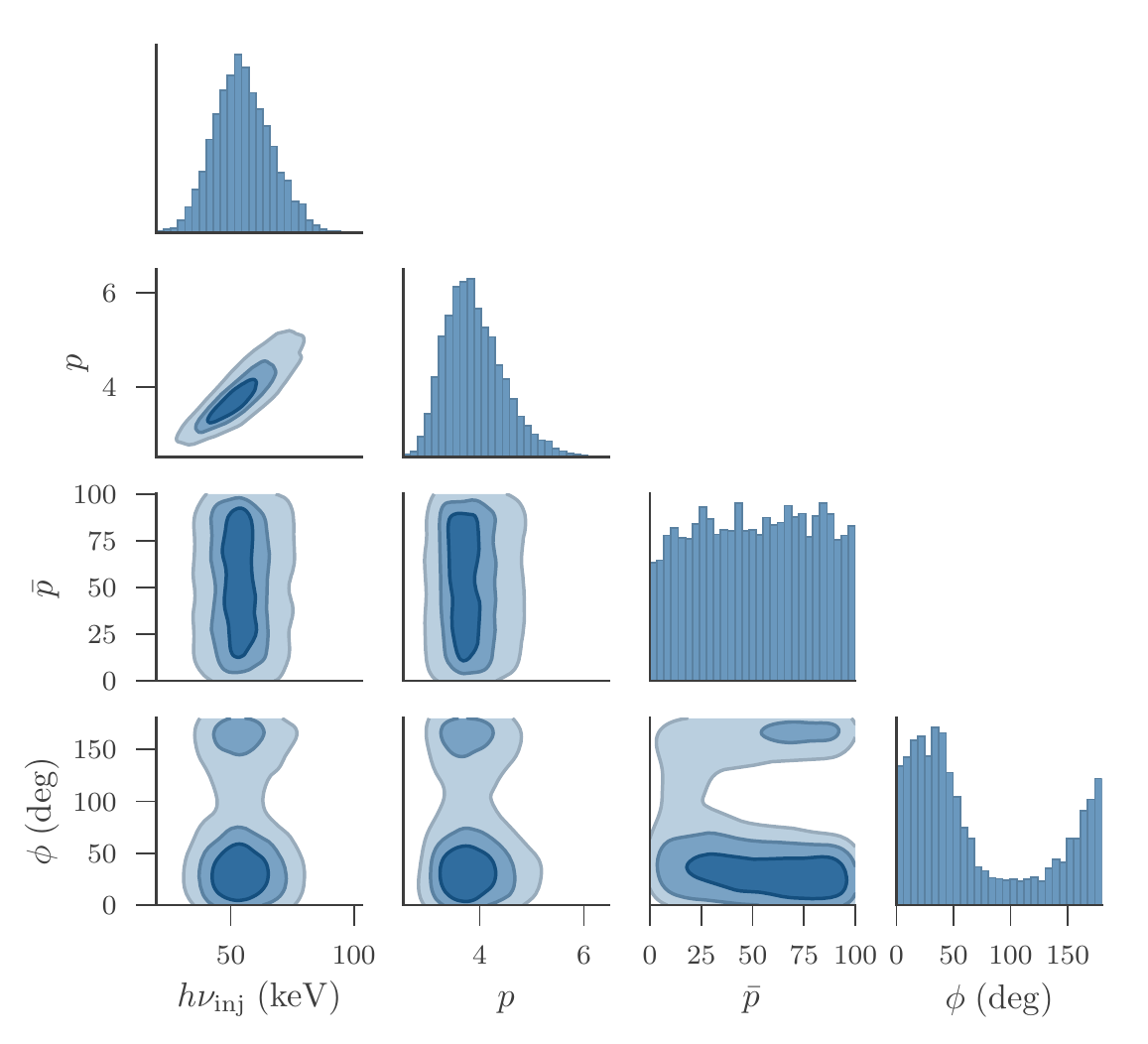}
  \caption{Parameter marginal distributions for time interval T: 8.9-20.0 s. }
  \label{fig:corner9}
\end{figure*}

\begin{figure}
  \centering
  \includegraphics{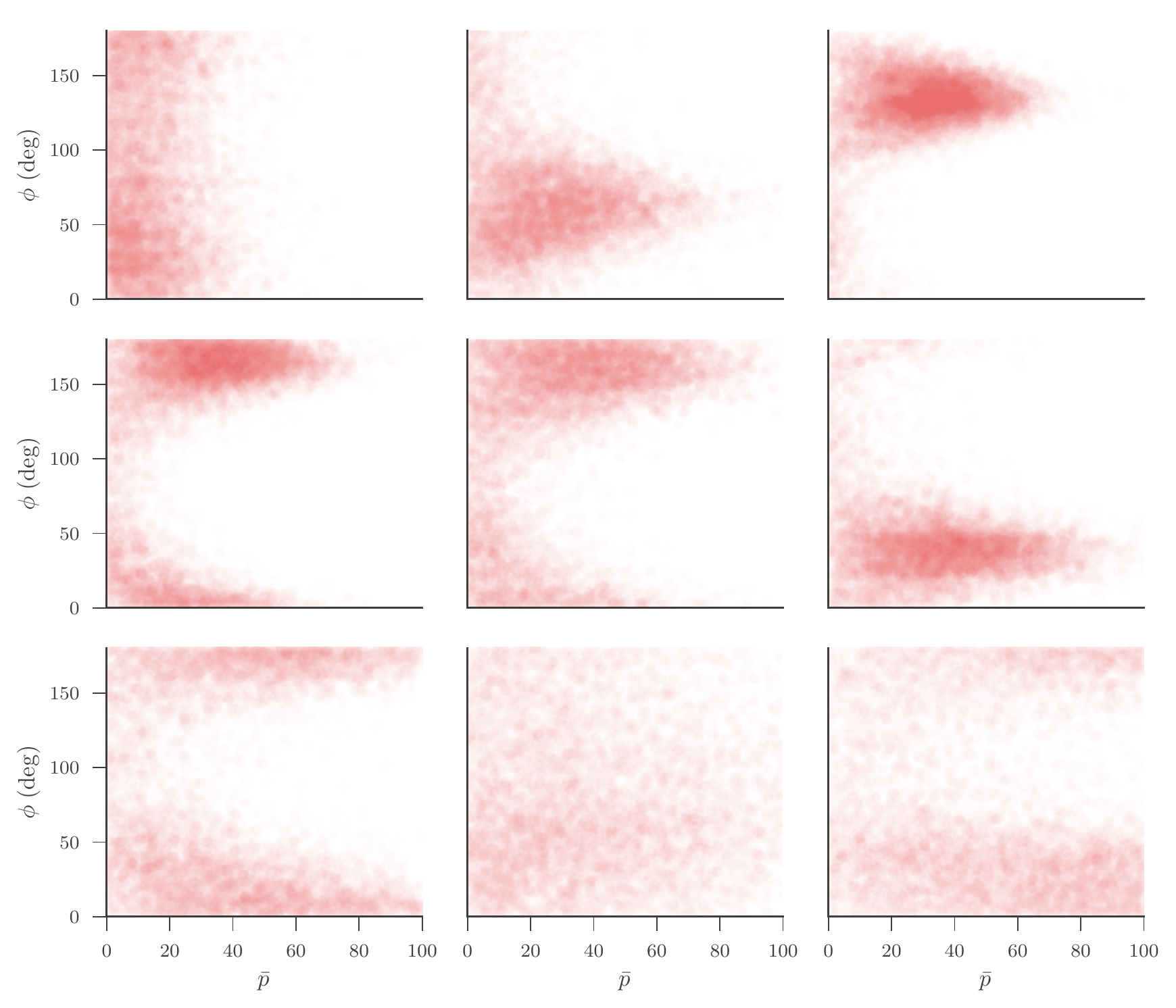}
  \caption{Posterior samples of the polarization quantities
    displayed in the common Cartesian projection. These samples
    correspond to those displayed in Figure \ref{fig:polarization} but
    are shown here for readers used to quantities displayed in this
    fashion.}
  \label{fig:pol_den}
\end{figure}

Here we present the parameter corner plots for the time intervals
described in the analysis sections. See Figures \ref{fig:corner1}
-\ref{fig:corner9} for these distributions. We have also plotted the
posterior samples from the polarization analysis on a Cartesian grid
for completeness in Figure \ref{fig:pol_den}.

\label{lastpage}

\end{document}